\documentclass[parskip=half]{scrartcl}

\usepackage{algorithm}
\usepackage{algpseudocode}
\usepackage{amsmath}
\usepackage{amssymb}
\usepackage{amsthm}
\usepackage{booktabs}
\usepackage{color}
\usepackage{graphicx}
\usepackage[colorlinks,citecolor=blue,urlcolor=blue]{hyperref}
\usepackage{mathbbol}
\usepackage[numbers]{natbib}
\usepackage{subcaption}
\usepackage{tikz}
\usepackage{tikz-qtree}

\newtheorem{theorem}{Theorem}
\newtheorem{definition}{Definition}

\newtheorem{proposition}{Proposition}

\newtheorem{corollary}{Corollary}

\newcommand{\1}{\boldsymbol{1}}

\newcommand{\bm}{{\boldsymbol{m}}}
\newcommand{\ba}{{\boldsymbol{a}}}
\newcommand{\by}{{\boldsymbol{y}}}

\newcommand{\bz}{{\boldsymbol{z}}}

\newcommand{\bw}{{\boldsymbol{w}}}

\newcommand{\bbb}{{\boldsymbol{\beta}}}
\newcommand{\beeta}{{\boldsymbol{\eta}}}
\newcommand{\bg}{{\boldsymbol{\gamma}}}

\newcommand{\bphi}{{\boldsymbol{\phi}}}
\newcommand{\bpsi}{{\boldsymbol{\psi}}}
\newcommand{\bmu}{{\boldsymbol{\mu}}}
\newcommand{\bxi}{{\boldsymbol{\xi}}}

\newcommand{\bsig}{{\boldsymbol{\sigma}}}

\newcommand{\bS}{{\boldsymbol{\Sigma}}}
\newcommand{\bT}{{\boldsymbol{T}}}
\newcommand{\bR}{{\boldsymbol{R}}}
\newcommand{\bL}{{\boldsymbol{L}}}
\newcommand{\bA}{{\boldsymbol{A}}}

\newcommand{\bQ}{{\boldsymbol{Q}}}
\newcommand{\bC}{{\boldsymbol{C}}}
\newcommand{\bB}{{\boldsymbol{B}}}

\newcommand{\bI}{{\boldsymbol{I}}}
\newcommand{\bX}{{\boldsymbol{X}}}
\newcommand{\bV}{{\boldsymbol{V}}}
\newcommand{\bU}{{\boldsymbol{U}}}
\newcommand{\bO}{{\boldsymbol{0}}}

\newcommand{\bigo}[1]{{\operatorname{\mathcal{O}}\left(#1\right)}}
\newcommand{\law}{\mathcal{L}}

\DeclareMathOperator{\Cost}{Cost}

\DeclareMathOperator{\Nor}{\mathcal{N}}

\DeclareMathOperator{\Cat}{Categorical}
\DeclareMathOperator{\Bin}{Binomial}
\DeclareMathOperator{\Wish}{Wishart}

\DeclareMathOperator{\Gam}{Gamma}

\DeclareMathOperator{\Cau}{Cauchy}
\DeclareMathOperator{\PG}{PG}
\DeclareMathOperator{\Unif}{Unif}
\DeclareMathOperator{\Pois}{Pois}

\begin{document}

\title{Scalable Bayesian computation for crossed and nested hierarchical models}
\author{Omiros Papaspiliopoulos \\ \href{mailto:omiros@unibocconi.it}{\texttt{[omiros@unibocconi.it]}} \and Timothee Stumpf-Fetizon \\ \href{mailto:timothee.stumpffetizon@unibocconi.it}{\texttt{[timothee.stumpffetizon@unibocconi.it]}} \and Giacomo Zanella \\ \href{mailto:giacomo.zanella@unibocconi.it}{\texttt{[giacomo.zanella@unibocconi.it]}}}
\maketitle

\begin{abstract}
We develop sampling algorithms to fit Bayesian hierarchical models, the computational complexity of which scales linearly with the number of observations and the number of parameters in the model.
We focus on crossed random effect and nested multilevel models, which are used ubiquitously in applied sciences.
The posterior dependence in both classes is sparse: in crossed random effects models it resembles a random graph, whereas in nested multilevel models it is tree-structured.
For each class we identify a framework for scalable computation, building on previous work.
Methods for crossed models are based on extensions of appropriately designed collapsed Gibbs samplers, where we introduce the idea of \emph{local centering}; while methods for nested models are based on sparse linear algebra and data augmentation.
We provide a theoretical analysis of the proposed algorithms in some simplified settings, including a comparison with previously proposed methodologies and an average-case analysis based on random graph theory.
Numerical experiments, including two challenging real data analyses on predicting electoral results and real estate prices, compare with off-the-shelf Hamiltonian Monte Carlo, displaying drastic improvement in performance.
The code for replicating the experiments in the article and for implementing our methods can be found at \url{https://github.com/timsf/crossed-effects} and \url{https://github.com/timsf/nested-effects}.
\end{abstract}

\section{Motivation and objectives}
\label{sec:intro}

In this article we explore efficient computational frameworks to perform Bayesian inference in large-scale hierarchical models. We focus on two canonical hierarchical structures, which are arguably the two most common in applied Statistics; crossed effects models, where the data is organized in a multidimensional contingency table and the within-cell distribution is modeled in terms of a linear expansion of a priori independent effects; and nested multilevel models, where data is organized in hierarchical clusters and parameters in a given level of the hierarchy are shrunk to (fewer) parameters in the deeper level. Crossed effects models are the canonical framework for modeling the dependence of an output variable on a number of categorical input variables. In the literature they appear under various names, e.g. cross-classified data, variance component models or multiway analysis of variance \cite{gelman2005analysis,searle2009variance,volfovsky2014hierarchical}. Nested multilevel models are the basic hierarchical structure for data and parameters that are naturally organized in nested clusters \cite{gelman2006data}. The detailed specifications of these models are given in Sections \ref{sec:crossed} and \ref{sec:nested}, respectively.

The number of parameters in these models will be denoted by $p$ and the number of observations by $n$. The computational challenge is to sample from the corresponding posterior distributions in high-dimensional scenarios where both $n$ and $p$ are large. The ``holy grail''  is MCMC algorithms whose total computational cost  scales linearly in $\max\{n,p\}$. We define carefully the notion of computational cost later in the article that involves both the cost per iteration and the number of iterations needed. We call algorithms whose cost is linear in  $\max\{n,p\}$, \emph{scalable}.

Both structures, crossed and nested, involve a high-dimensional vector of regression parameters, and a low-dimensional vector of hyperparameters. Priors for the regression parameters are usually Gaussian while the data likelihood is generic, and in this article we consider methods and numerics for a wide variety of such choices. The class of sampling algorithms we consider are (variations or approximations of) block-updating coordinate-wise schemes - a.k.a. \emph{blocked Gibbs samplers} - that alternate updating the regression parameters given the hyperparameters, and vice-versa. The main computational challenge, which is the focus of this paper, is how to update the high-dimensional regression parameters in a scalable way. 

The paper is structured as follows. Section \ref{sec:crossed} is devoted to crossed effect models. After describing the models and the high-level computational approach, in Section \ref{sec:gibbs_method_nGauss} we propose a methodology for scalable sampling of the high-dimensional regression parameters in crossed models with generic likelihood, using what we call a \emph{local centering} approach. In Section \ref{sec:crossed_multi}, we consider the practically important special case of multicategorical logistic likelihood where, beyond applying local centering ideas, we introduce a novel scheme for sampling efficiently without imposing identifiability constraints. These can instead be imposed, if desired, at a post-sampling stage. In Section \ref{sec:gibbs_polya}, in the specific context of  binomial likelihoods, we introduce an alternative scheme that combines the collapsed Gibbs sampler recently proposed in \cite{papaspiliopoulos2020scalable} with Polya-Gamma data augmentation \cite{polson2013bayesian}. Finally, Section \ref{sec:gibbs_complex} provides a theoretical analysis of the complexity of the above schemes, we establish scalability under certain assumptions and provide comparison results. In Section \ref{sec:nested} we consider nested multilevel models. We first show, in Section \ref{sec:gauss-bp}, how to use sparse linear algebra methods to update the high-dimensional regression coefficients for Gaussian likelihoods in a scalable way (which is not directly possible for crossed models). Then, in Section \ref{sec:ngauss-bp}, we combine the this methodology with data augmentation approaches to obtain scalable samplers for non-Gaussian likelihoods, such as binomial, Cauchy or Poisson.

The article contains various numerical experiments, including both simulated data and illustrations on two challenging large-scale applications, one on electoral survey data and one  on real-estate modeling. In all cases, we compare our proposed methods to state of the art alternatives, including Hamiltonian Monte Carlo, showing superior performance, often by more than an order of magnitude. Our numerical experiments also suggest that the conclusions from our theory extrapolate beyond the specific assumptions required for the theoretical analysis. Throughout the article, we use boldface letters to denote vectors (when lower case) and matrices (when upper case).

\section{Crossed effects models}
\label{sec:crossed}

We consider the following class of crossed effects models. There are $K$ categorical input variables, known as factors, each with $p_{k}$ categories, known as levels, and their numbering is arbitrary. In this context, $p = \sum_{k=1}^{K} p_{k}$ corresponds to the sum of levels over all factors. The combination of the levels of the different factors defines a $p_{1} \times \dots \times p_{k}$ contingency table. For each observation $j \in \{1, \dots, n\}$ we record the values of the categorical input variables and an output variable that could be multivariate and will be denoted by $\by_{j}$, with $L$ denoting the dimension of the response. Typical examples for $L=1$ are continuous, binary or count observations $y_{j}$, and multicategorical observations $\by_{j}$ for $ L > 1$, with details following below. The distribution of $\by_j$ depends on a linear predictor $\beeta_{j}$:
\begin{equation*}
 \beeta_{j}  = \bX_j \bbb + \ba^{(0)} + \textstyle\sum_{k=1}^{K} \ba_{i_{k}[j]}^{(k)}
\end{equation*}
where $\bX_{j}$ are continuous input variables (excluding an intercept), $\bbb$ and $\ba^{(0)}$ are what we typically call fixed effects and associated to each level of each categorical factor there is a random effect. The notation we use, which is quite standard in some of the relevant literature (see, e.g., Section 1.1 and Chapter 11 of \cite{gelman2006data}) is that $i_{k}[j]$ refers to the level that the $j$th observation has on the $k$th factor. Therefore, the available data per ``individual'' $j$ (in our election survey example, a survey respondent) are the output data $\by_{j}$ (e.g., a 1-hot encoded vector for the party they plan to vote for), the strata $i_k[j]$ they belong on each categorical factor $k$ (e.g., reside in Catalunya, unemployed, higher education etc), and other, if any, continuous inputs $\bX_{j}$. In our notation, $\ba_{i}^{(k)}$ is the random effect associated to the level $i$ of the $k$th factor, and its dimension is $L$, just as for the observations. We model all random effects as a priori independent with factor-dependent variances. The main reason why we allow for multivariate random effects in this article is to deal with multi-categorical responses, as in the election survey application, in which case there are precision matrices describing the dependence of the elements of each multivariate random effect. 

At this level of generality, with $\law$ denoting laws of random variables, the crossed effects model is as follows:
\begin{equation}\label{eq:anova}
\begin{aligned}
  \law(\ba^{(0)}) & = \Nor(\bmu_{pr}, \bT_{pr}^{-1}), \\ 
  \law(\bT_{k}) & = \Wish(\nu_{k}, \bI / \nu_{k}), \\
  \law(\ba^{(k)} \mid \bT_{k}) & = \Nor(\bO, (\bT_{k} \otimes \bI)^{-1} ),&k = 1,\ldots,K, \\
  \law(\by_{j} \mid \cdot) & = \law( \by_{j} \mid \beeta_{j}),\\
  \beeta_{j} & = \bX_j \bbb + \ba^{(0)} + \textstyle\sum_{k=1}^{K} \ba_{i_{k}[j]}^{(k)},  &j = 1,\ldots,n, \\
\end{aligned}
\end{equation}
where $\Nor$ denotes Gaussian distributions, $\law(\by_{j} \mid \cdot)$ denotes the law of $\by_{j}$ given all other variables, $\otimes$ stands for the Kronecker product, $\bT_{k}$ are $L \times L$ precision matrices,  $\ba^{(k)}$ is the vector of all stacked $\ba_{i}^{(k)}$'s, and the $\bI$'s are identity matrices of appropriate dimension. We understand the limiting case $\bmu_{pr} = \bO, \bT_{pr} = \bO \bO^{\mathrm{T}}$ as the improper flat prior on $\ba^{(0)}$. When $L=1$ we denote the corresponding precision scalars by $\tau_{k}$ and assign them gamma priors. The ``fixed effects'' coefficients $\bbb$ are typically assigned a Gaussian or flat prior.

In our numerical illustrations we focus on three important special cases: Gaussian models, where $L=1$ and $\law(y_{j} \mid \eta_{j},\tau) = \Nor(\eta_{j} , 1/\tau)$ (with a gamma prior on $\tau$); binomial models, where $L=1$ and $\law(y_{j} \mid \eta_{j}) = \Bin(m_{j}, (1+\exp\{-\eta_{j}\})^{-1})$; and multicategorical logit models, where $L > 1$, $\by_{j}$ is an $L$-dimensional vector of zeros and a single one (one-hot encoding), and $\law(\by_{j} \mid \beeta_{j}) = \Cat(\operatorname{softmax} \{\beeta_{j}\})$, with
\begin{equation*}\label{eq:softmax}
  \operatorname{softmax}\{\beeta\} = \left(\frac{e^{\eta_1}}{\sum_{l} e^{\eta_{l}}}, \cdots, \frac{e^{\eta_{L}}}{\sum_{l} e^{\eta_{l}}}\right).
\end{equation*}
The multicategorical logistic model introduces the further complication of being ill-identified due to softmax being invariant under translations of $\beeta$. In Section \ref{sec:crossed_multi} we address this and other methodological issues specific to this model.

\subsection{Computational strategy}
\label{sec:gibbs:intro}

For simplicity we focus on the case of no fixed effects ($\bbb=\bO$), but our methodology extends easily to when the model includes fixed effects, as we discuss below. In this context the regression-type parameters are $\ba = (\ba^{(0)}, \ba^{(1)}, \dots, \ba^{(K)})$. We let  $\bg$ refer collectively to the precision matrices $\bT_{k}$ and potential parameters specific to the likelihood $\law(\by_{j} \mid \beeta_{j})$. The methodology developed in the following sections is about sampling from $\law(\ba \mid \by, \bg)$, where $\by = (\by_{j})_{j=1,\dots,n}$ denotes all the observed data. We augment such methods with steps that update $\law(\bg \mid \by,\ba)$, which in our context is easily done by sampling from gamma and (low-dimensional) Wishart distributions to obtain a block coordinate-wise sampling algorithm for $\law(\ba, \bg \mid \by)$. Potential issues related to dependence between $\ba$ and $\bg$ in the joint posterior distribution $\law(\ba, \bg \mid \by)$ are discussed in the numerical parts and in the discussion on future work in Section \ref{sec:beyond}.

The methods we discuss are generally based on partitioning the regression parameters into $K+1$ blocks, $(\ba^{(0)}, \ba^{(1)}, \dots, \ba^{(K)})$. Previous work by \cite{gao2017efficient,papaspiliopoulos2020scalable} has proved that the complexity of the vanilla Gibbs sampler that updates each $\ba^{(k)}$ conditionally on the rest ($k=0,1,\ldots,K$) is typically super-linear in $n$ and $p$. \cite{papaspiliopoulos2020scalable} developed an approach based on a joint update of $(\ba^{(0)}, \ba^{(k)})$ for each $k=1,\ldots,K$, which under certain conditions is provably scalable. Their methodology, which they call collapsed Gibbs sampling, is only applicable for Gaussian likelihoods. Below we develop methodologies that extend these ideas to the case of general likelihoods. In some cases one can take advantage of data augmentation techniques, such as Polya-gamma data augmentation for binomial likelihoods, to restore conditional Gaussianity and directly apply the collapsed Gibbs sampler. We explore this in Section \ref{sec:gibbs_polya}. On ther other hand, in Section \ref{sec:gibbs_method_nGauss} we develop a methodology, called \emph{local centering}, that works with any (sufficiently smooth) likelihood, while inheriting the properties of the collapsed Gibbs sampler both in terms of convergence speed and cost per iteration. Note that, for Gaussian likelihoods, one could also perform a direct block update of the high-dimensional multivariate Gaussian distribution $\law(\ba \mid \by, \bg)$. However, the cost of this operation would be super-linear in $\max\{n,p\}$ for crossed effect models and thus not scalable. See also Section \ref{sec:crossed_disc} and related discussions in \cite{gao2017efficient,menictas2019streamlined,papaspiliopoulos2020scalable,ghosh2020backfitting}.

Regarding the inclusion of fixed effects $\bbb$ in the model, we can adapt the computational approach we develop here in various ways. The simplest strategy would be to update the global coefficients $\bbb$ from their full conditional distribution through a Gibbs or Metropolis-within-Gibbs update, following the update of $\ba$ detailed below. Regardless of the specific implementation, the key aspect is that the dimensionality of $\bbb$ is much lower than the dimensionality of $\ba$ in our applications of interest. More robust strategies are possible, such as updating $\bbb$ jointly with the intercept $\ba^{(0)}$, but we defer the discussion of such strategies, as well as of the discussion of the case of high-dimensional $\bbb$, to future work.

\subsection{Methodology for generic non-Gaussian likelihoods: local centering}
\label{sec:gibbs_method_nGauss}

In the same spirit as the collapsed Gibbs sampler of \cite{papaspiliopoulos2020scalable} for Gaussian likelihoods, we develop a scheme for the joint update of $(\ba^{(0)}, \ba^{(k)})$ that is feasible for generic likelihoods. To put our proposal into perspective, we first note that off-the-shelf sampling methods will not be successful. When factor $k$ has a large number of levels, this will be a high-dimensional sampling problem. Additionally, there is   strong posterior dependence between $\ba^{(0)}$ and $\ba^{(k)}$, due to the constraint imposed by the data via the linear predictor. The prior precision does not capture this dependence in crossed effects models, since a priori all variables are independent. Thus one cannot use gradient-based samplers for latent Gaussian models that precondition using the prior precision, as for example in \cite{titsias2018auxiliary}. Moreover, even with hypothetically good preconditioning, the convergence time of standard gradient-based samplers targeting the high-dimensional joint distribution of $(\ba^{(0)}, \ba^{(k)})$ will typically grow with $p_{k}$ (see e.g.\ \cite{roberts1998optimal,dwivedi2019log,vogrinc2022optimal,wu2022minimax} and related literature for various analysis of the dimensionality dependence of gradient-based MCMC algorithms), resulting in algorithms with super-linear complexity w.r.t.\ $n$ and $p$ (see Section \ref{sec:gibbs_complex} for proper definitions on convergence).  

Instead, we explicitly leverage the sparse conditional independence structure of $\law(\ba^{(0)}, \ba^{(k)} \mid \by, \bg, \ba^{(-0,-k)})$, where $\ba^{(-0,-k)}$ is a short form of $(\ba^{(\ell)})_{\ell\neq 0,k}$. 
We propose to use local hierarchical centering within each block:
\begin{equation}\label{eq:local-c}
  (\ba^{(0)}, \ba^{(k)}) \rightarrow (\ba^{(0)}, \bxi^{(k)}), \quad \bxi_{i}^{(k)} = \ba^{(0)} + \ba_{i}^{(k)}.
\end{equation}
Working locally with a centered parameterisation is motivated by the general principles of parameterisation of hierarchical models: as discussed in \cite{gelfand1995efficient, papaspiliopoulos2007general}, the model correlation structure is such that the conditional posterior dependence is weaker between the centered $(\ba^{(0)}, \bxi^{(k)})$ than the non-centered $(\ba^{(0)}, \ba^{(k)})$ pair. Now, note that due to the hierarchical structure, we obtain that $\law(\ba^{(0)} \mid \by, \bg, \ba^{(-0,-k)}, \bxi^{(k)}) = \law(\ba^{(0)} \mid \bxi^{(k)})$. Additionally, the $\bxi_{i}^{(k)}$ are independent across $i$ conditionally on $\ba^{(0)}$ and the levels of the other factors, hence a high-dimensional sampling problem reduces to $p_{k}$ independent $L$-dimensional ones. With these observations made, we propose Algorithm \ref{alg:local-c}.
\begin{algorithm}
\begin{algorithmic}
  \For{$k = 1:K$}
  \State Reparameterise $(\ba^{(0)}, \ba^{(k)}) \to (\ba^{(0)}, \bxi^{(k)})$ according to \eqref{eq:local-c}
  \State Draw $\ba^{(0)}$ from
\begin{equation*}
  \law(\ba^{(0)} \mid \bxi^{(k)}) = \Nor\begin{pmatrix}(\bT_{pr} + p_{k} \bT_{k})^{-1} \left(\bT_{pr} \bmu_{pr} + \bT_{k} \sum_{i=1}^{p_{k}} \bxi_{i}^{(k)}\right), \\ (\bT_{pr} + p_{k} \bT_{k})^{-1}\end{pmatrix}
\end{equation*}
  \For{$i = 1:p_{k}$}
  \smallskip
  \State Draw $\bxi_{i}^{(k)}$ from a kernel that leaves $\law(\bxi_{i}^{(k)} \mid \by, \bg, \ba^{(-k)})$ invariant
  \EndFor
  \State Reparameterise $(\ba^{(0)}, \bxi^{(k)}) \to (\ba^{(0)}, \ba^{(k)})$ according to \eqref{eq:local-c}
  \EndFor
\end{algorithmic}
\caption{Gibbs sampler with local centering for non-Gaussian likelihoods
\label{alg:local-c}}
\end{algorithm}

We now detail a gradient-based scheme that leaves $\law(\bxi_{i}^{(k)} \mid \by, \bg, \ba^{(-k)})$ invariant. The scheme is most concisely described by using the symbol $f_{i}^{(k)}(\ba_{i}^{(k)})$ to refer to the log-likelihood of $\by$ as a function of $\ba_{i}^{(k)}$, holding other parameters fixed. When $L=1$, we use a second-order Metropolis-Hastings proposal that requires no tuning parameters and is easy to implement. The approach is based on a second-order expansion of the corresponding density around the current value and amounts to generating a proposal $\tilde{\xi}_{i}^{(k)} \sim \Nor(m_{i}^{(k)}(\xi_{i}^{(k)}), c_{i}^{(k)}(\xi_{i}^{(k)}))$, where
\begin{equation*}
\begin{aligned}
  m_{i}^{(k)}(\xi_{i}^{(k)}) & = c_{i}^{(k)}(\xi_{i}^{(k)}) (\partial f_{i}^{(k)}(\xi_{i}^{(k)} - a^{(0)}) + \tau_{k} a^{(0)} - \partial^{2} f_{i}^{(k)}(\xi_{i}^{(k)} - a^{(0)}) \xi_{i}^{(k)}), \\
  c_{i}^{(k)}(\xi_{i}^{(k)}) & = \left(\tau_{k} - \partial^{2} f_{i}^{(k)}(\xi_{i}^{(k)} - a^{(0)})\right)^{-1},
\end{aligned}
\end{equation*}
with $\tau_{k}$ denoting the prior precision of $a_{i}^{(k)}$ for $L=1$. We accept $\tilde{\xi}_{i}^{(k)}$ with probability
\begin{equation}
  1 \wedge \frac{f_{i}^{(k)}(\tilde{\xi}_{i}^{(k)} - a^{(0)})}{f_{i}^{(k)}(\xi_{i}^{(k)} - a^{(0)})} \frac{\Nor (\tilde{\xi}_{i}^{(k)}; a^{(0)}, \tau_{k}^{-1})}{\Nor(\xi_{i}^{(k)}; a ^{(0)}, \tau_{k}^{-1})} \frac{\Nor(\xi_{i}^{(k)}; m_{i}^{(k)}(\tilde{\xi}_{i}^{(k)}), c_{i}^{(k)}(\tilde{\xi}_{i}^{(k)}))}{\Nor(\tilde{\xi}_{i}^{(k)}; m_{i}^{(k)}(\xi_{i}^{(k)}), c_{i}^{(k)}(\xi_{i}^{(k)}))}.
\end{equation}
When $L>1$, we use the gradient-based Metropolis-Hastings sampler of \cite{titsias2018auxiliary}, which proposes $\tilde{\bxi}_{i}^{(k)} \sim \Nor(\boldsymbol{m}_{i}^{(k)}, \boldsymbol{D}_{i}^{(k)})$, with
\begin{equation}\label{eq:marginal_expr}
\begin{aligned}
  \boldsymbol{m}_{i}^{(k)} & = \boldsymbol{C}_{i}^{(k)} (\bxi_{i}^{(k)} / \delta_{i}^{(k)} + \nabla f_{i}^{(k)}(\bxi_{i}^{(k)} - \ba^{(0)}) + \bT_{k} \ba^{(0)}), \\
  \boldsymbol{D}_{i}^{(k)} & = \boldsymbol{C}_{i}^{(k)} + \left(\boldsymbol{C}_{i}^{(k)}\right)^{2} / \delta_{i}^{(k)}, \\
  \boldsymbol{C}_{i}^{(k)} & = \left(\bT_{k} + \boldsymbol{I} / \delta_{i}^{(k)}\right)^{-1},
\end{aligned}
\end{equation}
and accepting it with probability
\begin{equation}\label{eq:marginal_acc_rej}
  1 \wedge \frac{f_{i}^{(k)}(\tilde{\bxi}_{i}^{(k)} - \ba ^{(0)})}{f_{i}^{(k)}(\bxi_{i}^{(k)} - \ba^{(0)})} \frac{\Nor(\tilde{\bxi}_{i}^{(k)}; \ba ^{(0)}, \bT_{k}^{-1})}{\Nor(\bxi_{i}^{(k)}; \ba^{(0)}, \bT_{k}^{-1})} \frac{\Nor(\bxi_{i}^{(k)}; \boldsymbol{m}_{i}^{(k)}(\tilde{\bxi}_{i}^{(k)}), \boldsymbol{D}_{i}^{(k)})}{\Nor(\tilde{\bxi}_{i}^{(k)}; \boldsymbol{m}_{i}^{(k)}(\bxi_{i}^{(k)}), \boldsymbol{D}_{i}^{(k)})}.
\end{equation}
The scalars $\delta_{i}^{(k)}$ are level-specific step size parameters that we tune adaptively according to the Robbins-Monro procedure described in, e.g., Algorithm 4 of \cite{andrieu2008tutorial}, with target acceptance rate equal to $0.5$ and learning rate at iteration $t$ set to $t^{-0.5}$. Notice that the matrices $\boldsymbol{D}_{i}^{(k)}$ for different $i$'s share the same eigenspace. Accordingly, an efficient implementation first eigendecomposes $\bT_{k} = \bV \boldsymbol{\Lambda} \bV^{\mathrm{T}}$, then computes
\begin{equation*}
  \sqrt{\boldsymbol{D}_{i}^{(k)}} = \bV \sqrt{(\boldsymbol{\Lambda} + \boldsymbol{I} / \delta_{i}^{(k)})^{-2} / \delta_{i}^{(k)} + (\boldsymbol{\Lambda} + \boldsymbol{I} / \delta_{i}^{(k)})^{-1}},
\end{equation*}
then simulates the proposal and evaluates its density.

Note that, while the local parameterisation proposed above is effective (as we show below), a global reparameterisation of the model would not be successful. The intuition is that one cannot simultaneously center all $K$ factors. Indeed, \cite[Theorem 5]{zanella2020multilevel} show that for crossed effects models with $K=2$ factors, all global reparameterisations based on hierarchical centering lead to a complexity that is super-linear in $n$ and $p$.

\begin{figure}[h]
  \centering
  \includegraphics{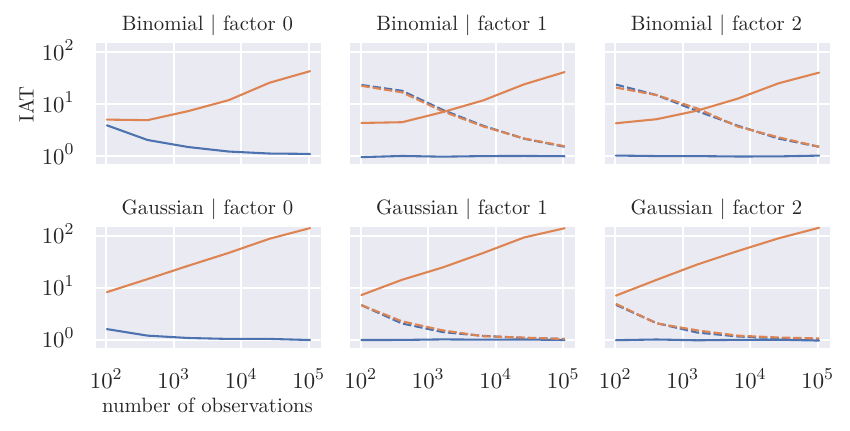}
  \caption{[\emph{Crossed effects simulation study, binomial and Gaussian likelihoods}] Integrated autocorrelation time estimates as a function of $n$ for the locally centered (blue) and the vanilla (orange) Metropolis-within-Gibbs sampler. Solid lines refer to $a^{(0)}$ and $p_{k}^{-1} \sum_{i=1}^{p_{k}} a_{i}^{(k)}$, and dashed to $\tau_{k}^{-1}$, for $k = 1,2$. Estimates are computed by averaging over 10 MCMC runs with 10000 iterations each. We set the priors $\tau_{k} \sim \Gam(1/2, 1/2)$, a flat prior on $a^{(0)}$ and $\tau \sim \Gam(1/2, 1/2)$ for the Gaussian model. Note that whereas the simulation of data was done with different values of $p_{k}$, here we show the results as a function of the implied $n$. \label{fig:crossed_lc_v_vanilla}}
\end{figure}

\subsubsection{Numerical illustration}

To demonstrate the scalability of the local centering approach, we generate data of growing dimension, and infer the corresponding models using the local centering and the vanilla Gibbs samplers. Elements of $\ba$ are sampled from a standard Gaussian with settings $L=1$, $K=2$, $p_{1} = p_{2}$ and increasing values of $p_{1}$. Observations $\by$ are missing completely at random, where each cell in the contingency table is missing  with probability 0.9, therefore $n \approx 0.1 p_{1}^{2}$. For the non-empty cells we simulate observations either from the Gaussian distribution $\law(y_{j} \mid \eta_{j}) = \Nor(\eta_{j}, 1)$, or from the binomial law $\law(y_{j} \mid \eta_{j}) = \Bin(1, (1 + \exp(-\eta_{j}))^{-1})$, and we use the corresponding likelihood when we learn the models. The precision parameters $\tau_{k}$ and the residual precision $\tau$ of the Gaussian model are treated as unknown.

We monitor the empirical estimates of \emph{integrated autocorrelation times} (IAT) for various parameters, and report them in Figure \ref{fig:crossed_lc_v_vanilla}. The IAT is a measure of MCMC sampling efficiency (less being better), whose definition and interpretation can be found in Appendix \ref{app:mcmc}. For the local centering algorithms this quantity remains upper bounded with respect to $n$ (and thus also of $p$). This happens for all parameters and suggests that the convergence properties of Algorithm \ref{alg:local-c} do not deteriorate as $n$ and $p$ increase. Combining this with considerations of cost per iteration, we conjecture that Algorithm \ref{alg:local-c} is scalable. Section \ref{sec:gibbs_complex} contains a more detailed discussion and some theoretical results that support the conjecture.

\subsection{Methodology for categorical likelihoods}
\label{sec:crossed_multi}

The local centering approach of Section \ref{sec:gibbs_method_nGauss} applies to any sufficiently smooth likelihood. A case of special interest, see e.g.\ the application to electoral survey data below, is the categorical logistic regression model with likelihood
\begin{equation}
  \law(\by_{j} \mid \beeta_{j}) = \Cat\left(\operatorname{softmax} \{\beeta_{j}\}\right)
\end{equation}
with $\operatorname{softmax}$ defined as in \eqref{eq:softmax}. In this model, $\law(\by_{j} \mid \beeta_{j}) = \law(\by_{j} \mid \beeta_{j} + c\1)$ for any constant $c$, where $\1$ denotes a vector of $1$'s of appropriate dimensionality. Thus, $\beeta_{j}$ is only identified by the likelihood up to such translations, which is also the case with the factor levels $\ba_{i}^{(k)}$. Of course, given proper priors, the resulting posterior is proper. Therefore, we set proper priors throughout this section, including on $\ba^{(0)}$, albeit weakly informative. Even with a proper prior, the posterior is typically much more informative about the differences in the elements of $\ba_{i}^{(k)}$ than it is about the mean of said vector. As a consequence, the elements have strong posterior dependence, and an MCMC algorithm would need to propose highly correlated moves to fully traverse the posterior. In particular, the Metropolis-Hastings algorithm introduced in Section \ref{sec:gibbs_method_nGauss} will fail to propose such moves, since it approximates $\law(\bxi_{i}^{(k)} \mid \by, \bg, \ba^{(0)}, \ba^{(-k)})$ by way of its gradient and the prior covariance, neither of which capture the extent of the dependence. Therefore, its mixing for categorical likelihoods can be poor, and we provide some numerical evidence to that effect in Figure \ref{fig:crossed_multinomial_ident}.

Since the differences of the elements of $\ba_{i}^{(k)}$ \emph{are} usually well-identified, we do expect the Metropolis-Hastings algorithm to mix adequately once the magnitude of the vector has been fixed. Therefore, one way of addressing this issue is to change the prior on the factor levels by imposing an identifiability constraint. One such constrain is to set the $L$-th element $a_{iL}^{(k)}$ of the vector $\ba_{i}^{(k)}$ to 0, define the remaining $L-1$ elements as $\dot{\ba}_{i}$, and specify the Gaussian prior
\begin{equation}
  \law(\dot{\ba}_{i}^{(k)} | \dot{\bT}_{k}) = \Nor(\bO, \dot{\bT}_{k}^{-1}),
\end{equation}
with an analogous constraint on $\ba^{(0)}$. We may then proceed as in Section \ref{sec:gibbs_method_nGauss}, replacing the symbols $(\ba^{(0)}, \ba^{(k)}, \bT_{0}, \bT_{k})$ with $(\dot{\ba}^{(0)}, \dot{\ba}^{(k)}, \dot{\bT}_{0}, \dot{\bT}_{k})$. In doing so, we keep in mind that since $a_{iL}^{(k)} = 0$, the linear predictors have a trailing element fixed to 0, i.e.
\begin{equation}\label{eq:linpredident}
  \beeta_{j} = \ba^{(0)} + \sum_{k=1}^{K} \ba_{i_{k}[j]}^{(k)} = \begin{bmatrix} \dot{\ba}^{(0)} + \sum_{k=1}^{K} \dot{\ba}_{i_{k}[j]}^{(k)} \\ 0 \end{bmatrix},
\end{equation}
which needs to be accounted for when evaluating the likelihood and its gradient.

While this solution is computationally effective, it is statistically unsatisfactory because different identifiability constraints lead to different posterior inferences for the outcome probabilities of the categorical variable. We think it preferable to improve the algorithm without changing the model, and leave it to the modeler to impose constraints post-hoc if desired. Hence, we propose a novel scheme that, although conceptually close to the identifiability constraint described above, does not change the prior but rather changes the sampling algorithm. Rather than tethering one of the elements to 0, we follow the principle set out in Section \ref{sec:gibbs_method_nGauss} by updating the well-identified quantities by Metropolis-within-Gibbs, and the ill-identified translation through an additional Gibbs step which only involves the prior. To obtain the well-identified quantity, define $\ddot{\ba}_{i}^{(k)}$ as the first $L-1$ elements of $\ba_{i}^{(k)} - a_{iL}^{(k)} \1$, and similarly $\ddot{\ba}^{(0)}$ as the first $L-1$ elements of $\ba^{(0)} - a_{L}^{(0)} \1$. Note that the data $\by$ are conditionally independent of $a_{iL}^{(k)}$ and $a_{L}^{(0)}$ given $(\ddot{\ba}_{i}^{(k)}, \ddot{\ba}^{(0)})$. Additionally, if $\law(\ba_{i}^{(k)} | \bg) = \Nor(\bO, \bT_{k}^{-1})$ with
\begin{equation}
  \bT_{k}^{-1} = \begin{bmatrix} \bS_{(-L,-L)}^{(k)} & \bsig_{(-L, L)}^{(k)} \\ \bsig_{(L, -L)}^{(k)} & \sigma_{(L, L)}^{(k)} \end{bmatrix},
\end{equation}
then the implied prior on $\ddot{\ba}_{i}^{(k)}$ is $\law(\ddot{\ba}_{j}^{(k)} | \bg) = \Nor(\bO, \ddot{\bT}_{k}^{-1})$, where
\begin{equation}
  \ddot{\bT}_{k}^{-1} = \bS_{(-L,-L)}^{(k)} + \sigma_{(L, L)}^{(k)} \1\1^{\mathrm{T}}  - \1 \bsig_{(L, -L)}^{(k)} - \bsig_{(-L, L)}^{(k)} \1^{\mathrm{T}},
\end{equation}
with corresponding expressions for $\ddot{\ba}^{(0)}$. Thus, our proposed scheme is to apply the methodology of Section \ref{sec:gibbs_method_nGauss} starting from $(\ddot{\ba}_{i}^{(k)}, \ddot{\ba}^{(0)})$ and their implied priors, merely replacing $(\ba^{(0)}, \ba^{(k)}, \bT_{0}, \bT_{k})$ with $(\ddot{\ba}^{(0)}, \ddot{\ba}^{(k)}, \ddot{\bT}_{0}, \ddot{\bT}_{k})$, and again keeping in mind that the linear predictors are as in \eqref{eq:linpredident}. We then recover $a_{iL}^{(k)}$ by observing that
\begin{equation}
\begin{aligned}
  \law(a_{iL}^{(k)} | \bg, \ddot{\ba}_{i}^{(k)}) & = \Nor(\boldsymbol{b}_{k}^{\mathrm{T}} \ddot{\ba}_{i}^{(k)}, \boldsymbol{b}_{k}^{\mathrm{T}} (\bsig_{(-L, L)}^{(k)} - \sigma_{L, L}^{(k)} \1)), \\
  \boldsymbol{b}_{k} & = \ddot{\bT}_{k} (\bsig_{(-L, L)}^{(k)} - \sigma_{L, L}^{(k)} \1),
\end{aligned}
\end{equation}
with similar expressions for $a_{L}^{(0)}$. Thereafter, we transform back according to
\begin{equation*}
  \left(\begin{bmatrix} \ddot{\ba}_{i}^{(k)} + a_{iL}^{(k)} \1 \\ a_{iL}^{(k)} \end{bmatrix}, \begin{bmatrix} \ddot{\ba}^{(0)} + a_{L}^{(0)} \1 \\ a_{L}^{(0)} \end{bmatrix}\right) \to (\ba_{i}^{(k)}, \ba^{(0)}).
\end{equation*}
In effect, the resampling of $a_{L}^{(0)}$ according to simple prior-induced Gaussian computations induces the required positive dependence in the update. This procedure is repeated for all factors. We refer to an algorithm using the above procedure as a \emph{projected} algorithm. The numerical experiments of the following section suggest that the projected algorithm with local centering mixes at least as well as its counterpart based on prior constraints. If required, such constraints can still be imposed post-sampling.

\subsubsection{Illustration on electoral survey data}

We experimentally assess algorithmic performance on a large-scale electoral survey dataset. The former previously featured in \cite{montalvo2019bayesian}, and allows us to assess algorithms under realistic conditions that arise in applied work. Electoral surveys, consisting of the self-reported voting intention and various demographic traits for a sample of potential voters, provide one of the most natural applications for crossed-effects modeling, see for example \cite{montalvo2019bayesian, goplerud2022fast} and references therein. Here the response variable is the multicategorical voting intention, with each category corresponding to a different party. Demographic traits such as location and age serve as input variables; these may be stratified into multicategorical variables. Electoral surveys are useful both for disaggregating the national voting intention to local districts, which is crucial for predicting parliamentary representation, and for dealing with insurgent political parties with significant electoral success, which is becoming a common phenomenon in European politics. The survey was carried out by the \emph{Centro de Investigaciones Sociol\'ogicas} ahead of the November 2019 Spanish general elections. For the experiments in this paper we focus on a simplified data structure with $L=3$ parties, $K=7$ input variables, and $210$ regression parameters overall (3 per level, and $p=70$ levels overall). There are $n=9234$ survey respondents and the input variables define a contingency table with  $33696$ cells, of which only 4764 are non-empty, hence the table is \emph{sparse}. Table \ref{tb:survey} summarises the dataset characteristics.

\begin{table}
\centering
\begin{tabular}{rlr}
\toprule
$k$ & factor name & no. of levels ($p_{k}$) \\
\midrule
1 & \ttfamily{province id} & 52 \\
2 & \ttfamily{activity} & 4 \\
3 & \ttfamily{age} & 3 \\
4 & \ttfamily{education} & 3 \\
5 & \ttfamily{municipality size} & 3 \\
6 & \ttfamily{last vote} & 3 \\
7 & \ttfamily{gender} & 2 \\
\midrule
- & response & 3 \\
\bottomrule
\end{tabular}
\caption{Categorical factors included in the crossed effects elections model. \label{tb:survey}}
\end{table}

The response being multi-categorical, we find ourselves in the setting of Section \ref{sec:crossed_multi}. In view of that discussion, we compare four algorithms: the vanilla Metropolis-within-Gibbs sampler for the constrained model (const/Van-MwG), the locally centered algorithm for the constrained model (const/LC-MwG), the locally centered algorithm for the unconstrained model (free/LC-MwG), and the projected local centering algorithm for the unconstrained model (free/PLC-MwG). Table \ref{tb:acro} in the Appendix provides a list of the algorithms discussed in this article and their acronyms. Notice that the first two sample from a different posterior than the last two.  

\begin{figure}[h]
  \centering
  \includegraphics{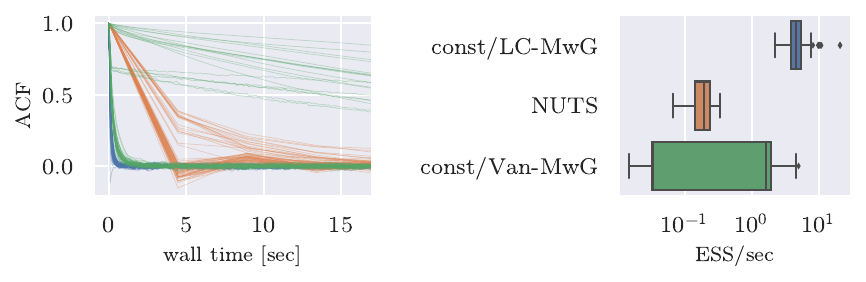}
  \caption{[\emph{Crossed effects elections model, constrained categorical likelihood}] Comparison of sampling efficiency between the locally centered sampler (const/LC-MwG, blue), the vanilla Metropolis-within-Gibbs sampler for the constrained model (const/Van-MwG) and Stan/NUTS (orange). The left panel overlays parameter ACFs of $\ba$ and $\bg$ as a function of wall time. The right panel shows the distribution of effective samples per second. Estimates are computed from MCMC runs with 10000 (const/LC-MwG, const/Van-MwG) and 1000 iterations (NUTS) respectively, with an identical number of iterations used for adaptation. We set the priors $\dot{\bT}_{k} \sim \Wish(L-1, \bI / (L-1))$, $\dot{\ba}^{(0)} \sim \Nor(\bO, \bI)$. \label{fig:crossed_multinomial_benchmark}}
\end{figure}

We carry out two experiments on this elections dataset, one of which is a comparison between const/Van-MwG, const/LC-MwG and the generic Stan implementation of NUTS, showing that const/LC-MwG provides a consistent speedup of more than an order of magnitude compared to Stan and const/Van-MwG, see Figure \ref{fig:crossed_multinomial_benchmark}. The const/Van-MwG has inconsistent performance, mixing fairly well in the factors $k$ with many levels, which identifies the mean of $\ba^{(k)}$ and breaks dependence with $\ba^{(0)}$. Factors with few levels remain strongly dependent with $\ba^{(0)}$ a posteriori, and therefore mix slowly. This finding is in accordance with the theoretical results of \cite{papaspiliopoulos2020scalable} (e.g., their Theorem 3) that predict this behaviour, albeit under conditions not met in this situation.

\begin{figure}[h]
  \centering
  \includegraphics{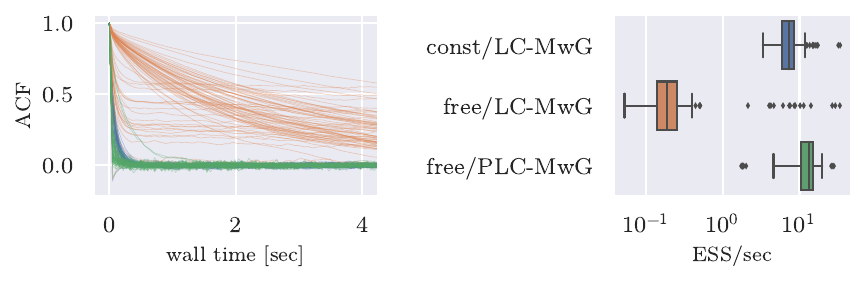}
  \caption{[\emph{Crossed effects election model, categorical likelihood}] Comparison of sampling efficiency between the locally centered sampler on the constrained model (const/LC-MwG, blue), the local centering sampler on the free model (free/LC-MwG, orange) and the projected local centering sampler on the free model (free/PLC-MwG, green). The left panel overlays parameter ACFs of $\ba$ and $\bg$ as a function of wall time. The right panel shows the distribution of effective samples per iteration. Estimates are computed from MCMC runs with 10000 iterations each, with an identical number of iterations used for adaptation. We set the priors $\bT_{k} \sim \Wish(L, \bI / L)$, $\ba^{(0)} \sim \Nor(\bO, \bI)$ in the free likelihood case and $\dot{\bT}_{k} \sim \Wish(L-1, \bI / (L-1))$, $\dot{\ba}^{(0)} \sim \Nor(\bO, \bI)$ in the constrained likelihood case. \label{fig:crossed_multinomial_ident}}
\end{figure}

The other experiment is a comparison between the various bespoke Gibbs samplers, which illustrates the importance of either constraining or projecting before local centering, see Figure \ref{fig:crossed_multinomial_ident}. Both the const/LC-MwG and the free/PLC-MwG algorithms perform well, while the mixing of most parameters in the free/LC-MwG algorithm is two orders of magnitude slower. The extent of the slowdown depends on the strength of the prior, which can restrict the possible range of translations in the softmax; $\ba^{(0)}$ and factors $k$ with many levels and hence well-identified hyperparameter $\bT_{k}$ mix better. Since const/LC-MwG and free/PLC-MwG lead to similar computational efficiency, we recommend the use of the unconstrained model with the  free/PLC-MwG algorithm and impose, if required, additional constraints in a post-sampling stage.

\subsection{Methodology for binomial likelihoods: Polya-gamma augmentation and collapsed Gibbs sampling}
\label{sec:gibbs_polya}

While the local centering approach described above has the appeal of applying to general likelihoods, it is not optimal for Gaussian likelihood since in that instance we can directly sample $\law(a^{(0)}, \ba^{(k)} \mid \by, \bg, \ba^{(-0,-k)})$ by exploiting Gaussian properties, as shown in \cite{papaspiliopoulos2020scalable} (and revisited later in this section). For certain non-Gaussian likelihoods we can still benefit from this direct approach when there are latent variables $\bz$ such that $\law(a^{(0)}, \ba^{(k)} \mid \by, \bz, \bg, \ba^{(-0,-k)})$ is Gaussian. We focus here on one such case of significant practical importance, that of binomial likelihod.

\cite{polson2013bayesian} observed that when $\law(y_{j} \mid \eta_{j}) = \Bin(m_{j}, (1 + e^{-\eta_{j}})^{-1})$, then the corresponding probability mass function admits the following mixture representation:
\begin{equation}
  \frac{(e^{\eta_{j}})^{y_{j}}}{(1 + e^{\eta_{j}})^{m_{j}}} = \int_{0}^{\infty} \frac{e^{(y_{j} - m_{j} / 2) \eta_{j} - z_{j} \eta_{j}^{2} / 2}}{2^{m_{j}}} \PG(z_{j}; m_{j}, 0) dz_{j}
\end{equation}
where $\PG(z; \alpha, \beta)$ denotes the density of the so-called \emph{Polya-gamma} distribution with parameters $\alpha$ and $\beta$ evaluated at $z$. The Polya-gamma distribution is constructed in \cite{polson2013bayesian} as a weighted, infinite sum of Gamma random variables. For integer-valued $\alpha$, a $\PG(\alpha, \beta)$-variable is obtained as a sum of $\alpha$ $\PG(1, \beta)$-variables, while simulation of a $\PG(1, \beta)$-variable is done by exploiting the representation of its density as a bounded Cauchy series whose increments have alternating signs.

This neat construction leads to the following computational strategy. We augment the space with the vector $\bz = (z_{1}, \ldots, z_{n})$ and design a sampler for $\law(\ba, \bg, \bz \mid \by)$. It is fairly simple to check (see \cite{polson2013bayesian}) that
\begin{equation}
 \law(\bz \mid \by, \ba, \bg) = \prod_{j=1}^{n} \PG(m_{j}, \eta_{j}),
\end{equation}
with $m_{j}$ and $\eta_{j}$ defined above. Another direct calculation shows that $\law(\ba \mid \by, \bz, \bg)$ is Gaussian. Then, we can follow the same paradigm as in Section \ref{sec:gibbs_method_nGauss} but sample  $\law(a^{(0)}, \ba^{(k)} \mid \by, \bz, \bg, \ba^{(-0,-k)})$ directly from the corresponding Gaussian distribution. Algorithm \ref{alg:collapsed} summarizes how this step is carried out.

\begin{algorithm}[h]
  Define the quantities
  \begin{equation}
    v_{j} = (y_{j} - m_{j} / 2) / z_{j}, \quad
    z_{i}^{(k)} = \textstyle\sum_{j: i_{k}[j] = i} z_{j}, \quad
    s_{i}^{(k)} = z_{i}^{(k)} / (\tau_{k}  + z_{i}^{(k)}).
  \end{equation}
\begin{algorithmic}
  \For{$k = 1:K$}
  \State Draw $a^{(0)}$ from
  \begin{equation*}
  \law(a^{(0)} \mid \by, \ba^{(-0,-k)}) = \Nor\begin{pmatrix}\frac{1}{\sum_{i} s_{i}^{(k)}} \sum_{i} \frac{s_{i}^{(k)}}{z_{i}^{(k)}} \sum_{j: i_{k}[j] = i} z_{j} (v_{j} - \sum_{k' \neq k} a_{i_{k'}[j]}), \\ (\tau_{k} \sum_{i} s_{i}^{(k)})^{-1}\end{pmatrix}
  \end{equation*}
  \For{$i = 1:p_{k}$}
  \smallskip
  \State Draw $a_{i}^{(k)}$ from
  \begin{equation*}
  \law(a_{i}^{(k)} \mid \cdot) = \Nor\begin{pmatrix} \frac{s_{i}^{(k)}}{z_{i}^{(k)}} \sum_{j: i_{k}[j] = i} z_{j} (v_{j} - a^{(0)} - \sum_{k' \neq k} a_{i_{k'}[j]}), \\ (z_{i}^{(k)} + \tau_{k})^{-1} \end{pmatrix}
  \end{equation*}
  \EndFor
  \EndFor
\end{algorithmic}
\caption{The collapsed Gibbs sampler for $\law(\ba| \by, \bg, \bz)$. \label{alg:collapsed}}
\end{algorithm}

An obvious advantage of this approach over the one based on local centering, when applicable, is the avoidance of the Metropolis-Hasting step, which typically slows down mixing and requires tuning. However, there are two disadvantages to the Polya-gamma augmentation. One is that the dimension of the distribution to be sampled is increased by the $n$ additional latent variables. Another is that sampling  $\PG(m_{j}, \eta_{j})$ has computational cost $\bigo{m_{j}}$, hence it will be inefficient for responses with large counts (although see \cite{windle2014sampling} for some mitigation of this problem).

Notice that we may contrast the above algorithm, which we call the Polya-Gamma collapsed Gibbs sampler (PG/Col-G), with another sampler on the augmented space that updates all effects one at a time and we call  the Polya-Gamma vanilla Gibbs sampler (PG/Van-G). We also compare those samplers to the  local centering algorithm (LC-MwG) applied to binomial likelihood.

\begin{figure}[h]
  \centering
  \includegraphics{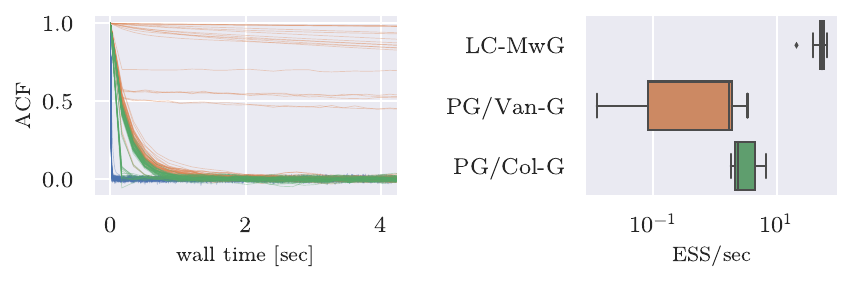}
  \caption{[\emph{Crossed effects election model, binomial likelihood}] Comparison of sampling efficiency between the collapsed Gibbs sampler on the augmented space (PG/Col-G, green), the locally centered marginal sampler (LC-MwG, blue) and the vanilla Gibbs sampler on the augmented space (PG/Van-G, orange). The left panel overlays parameter ACFs of $\ba$ and $\bg$ as a function of wall time. The right panel shows the distribution of effective samples per iteration. Estimates are computed from MCMC runs with 10000 iterations each. We set the priors $\tau_{k} \sim \Gam(1/2, 1/2)$ and a flat prior on $a^{(0)}$. \label{fig:crossed_binomial_elections}}
\end{figure}

Figure \ref{fig:crossed_binomial_elections} shows that the scalable algorithms LC-MwG and PG/Col-G significantly outperform the non-scalable algorithm PG/Van-G on the elections dataset. As in Figure \ref{fig:crossed_multinomial_benchmark}, the vanilla algorithm slows down the most for the factors with few levels. Note that while PG/Col-G and LC-MwG have similar ESS per iteration, PG/Col-G has substantially higher iteration time due to the latent variable update.

\subsection{Theoretical analysis}
\label{sec:gibbs_complex}

We now perform some theoretical analysis of the complexity of the algorithms described above. The computational complexity of an MCMC algorithm depends both on the cost per iteration and on the number of iterations required to obtain each effective sample. In this work, we quantify the number of iterations required through the relaxation time of the associated Markov chain (other options are possible, but we do not discuss them here), leading to the following definition.

\begin{definition}
We define the complexity of an MCMC algorithm as
\begin{equation}\label{eq:cost_gibbs_def}
  \Cost(\mathrm{MCMC}) = \text{(cost per iteration) $\times$ (relaxation time)},
\end{equation}
where the relaxation time refers to the reciprocal of 1 minus the geometric rate of convergence of the associated Markov chain \cite[Proposition 1]{rosenthal2003asymptotic}.
\end{definition}

We now analyse the cost per iteration and the relaxation time of the proposed schemes. Following the discussion in Sections \ref{sec:intro} and \ref{sec:gibbs:intro}, our theoretical analysis applies to the algorithm for sampling $\law(\ba \mid \by, \bg)$, even though all our numerics pertain to sampling $\law(\ba, \bg \mid \by)$.

\subsubsection{Cost per iteration of the proposed schemes}

The cost per iteration of the Gibbs sampler based on local centering proposed in Section \ref{sec:gibbs_method_nGauss} is of order $\bigo{nKL + KL^{3} + pL^{2}}$, as we show now. 
The evaluation of the gradients terms in \eqref{eq:marginal_expr} has a $\bigo{nL}$ cost for each factor, since
\begin{equation*}
  \nabla f_{i}^{(k)}(\ba_{i}^{(k)}) =
  \sum_{j:i_{k}[j]=i} \nabla_{\ba_{i}^{(k)}} \log p(\by_{j} \mid \ba),
\end{equation*}
implies that exactly $N$ terms of the form $\nabla_{\ba_{i}^{(k)}} \log p(\by_{j} \mid \ba)$ for $j = 1, \dots, n$ need to be computed, at $\bigo{L}$ cost each. By similar arguments also the likelihood evaluations required by the acceptance probability computation require $\bigo{nL}$ operations per factor. Given the gradient and likelihood computations, one can eigendecompose $\bT_{k}$ at $\bigo{L^{3}}$ cost and then perform the updates of $\bxi_{i}^{(k)}$ for all $i$ at $\bigo{p_{k}L^{2}}$ cost. Summing over factors we obtain the $\bigo{nKL + KL^{3} + pL^{2}}$ cost. The update from $\law(\ba^{(0)} \mid \bxi^{(k)})$ requires $\bigo{L^{3}}$ operations, which does not influence the overall cost and can further be reduced to $\bigo{L^{2}}$ if the prior precision $\bT_{pr}$ is diagonal and $\bT_{k}$ has already been eigendecomposed.

Conveniently, with $L=1$, this formula also holds for the collapsed Gibbs sampler for binomial likelihoods of Section \ref{sec:gibbs_polya}, i.e. the complexity of updating $\law(\ba^{(0)},\ba^{(k)} | \by, \ba^{(-0, -k)}, \bg, \bz)$ for all $k=1,\dots,K$ is $\bigo{nK + p}$. This is due to needing to sample from $p$ Gaussian distributions, and carrying out $\bigo{nK}$ summations to calculate their parameters.
When sampling $\law(\bz \mid \by, \ba,\bg)$ we incur the additional cost of $\bigo{\sum_{j} m_{j}}$, for a total cost of $\bigo{nK + p + \sum_{j} m_{j}}$.

For the categorical const/LC and free/pLC algorithms, notice that if the response is $L+1$-categorical, then local centering is done in $L$ dimensions. The additional computations in the free/pLC algorithm can also be carried out with complexity $\bigo{KL^{3}}$. Conversely, for free/pLC, the update to $\gamma$ is still $L+1$-dimensional, therefore the complexity of the full algorithm is $\bigo{nKL + K(L+1)^{3} + pL^{2}}$.

\subsubsection{Relaxation time analysis}

While the relaxation time of MCMC algorithms is hard to characterize in general, that of the Gibbs sampler for crossed effects models with Gaussian likelihood has become more amenable to explicit computations combining the classical results of \cite{roberts1997updating} with the so-called multigrid decomposition developed in \cite{zanella2020multilevel,papaspiliopoulos2020scalable}. In this section we extend those results to analyze and compare the sampling algorithms discussed in previous sections, providing two main results. The first, contained in Proposition \ref{prop:loc_c_averages}, Corollary \ref{thm:loc_c_equiv_to_collaps} and Theorem \ref{thm:loc_c_equiv_to_collaps}, provides an analysis of local centering and a comparison of the latter with the collapsed Gibbs Sampler. These results show that local centering always increases the relaxation time compared to exact collapsing, but does so only by a constant factor, thus leading to the same asymptotic complexity. The second contribution, formalized in Theorem \ref{thm:average_case}, is to expand the results in \cite{papaspiliopoulos2020scalable} by leveraging connections to random graph theory to show that both collapsing and local centering induce algorithms that are scalable as $n$ and $p$ diverge for average-case observation designs.

In this section we focus on models with Gaussian likelihood. Also, we assume $L=1$ to keep the notation simple, even if most of the results would naturally extend to the $L>1$ case. More precisely, we study algorithms targeting the law $\law(\ba \mid \by, \bg)$ induced by the following model
\begin{equation}\label{eq:crossed_Gauss1d}
\begin{aligned}
  \law(a^{(0)}) & = \Nor(\mu_{pr}, \tau_{pr}^{-1}), \\
  \law(\ba^{(k)} \mid \tau_{k}) & = \Nor(\bO, (\tau_{k} \bI)^{-1} ), & k = 1, \ldots, K \\
  \law(y_{j}|\ba,\bg) & = \Nor(a^{(0)} + \textstyle\sum_{k=1}^{K} a_{i_{k}[j]}^{(k)},\tau^{-1}) & j = 1, \ldots, n,\\
\end{aligned}
\end{equation}
for fixed hyperparameters $\bg=(\tau,\tau_{1},\dots,\tau_{k})$.
We compare the following three schemes:

\begin{enumerate}
\item[(GS)] the vanilla Gibbs sampler that updates from $\law(\ba^{(k)} \mid \by, \bg, \ba^{(-k)})$ for $k = 0, \dots, K$;
\item[(cGS)] the collapsed Gibbs sampler updating from $\law(a^{(0)}, \ba^{(k)} \mid \by, \bg, \ba^{(-0,-k)})$ for $k = 1, \dots, K$;
\item[(Loc)] the Gibbs sampler based on local centering that executes Algorithm \ref{alg:local-c} with the update of $\xi_{i}^{(k)}$ being the exact update from the conditional $\law(\xi_{i}^{(k)} \mid \by, \bg, \ba^{(-k)})$, which is feasible in the Gaussian case.
\end{enumerate}

We first define some concepts and quantities related to the data observation pattern.
\begin{definition}\label{def:cooc}
 We denote occurrence and co-occurrence counts associated to levels of different factors by
\begin{equation*}
\begin{aligned}
  n_{i}^{(k)} & = \sum_{j=1}^{n} \1(i_{k}[j] = i), \\
  n_{ii'}^{(k, k')} & = \sum_{j=1}^{n} \1(i_{k}[j] = i, i_{k'}[j] = i').
\end{aligned}
\end{equation*}
We say that a design has balanced levels if $n_{i}^{(k)} = n / p_{k}$ for each factor $k$ and level $i$. We define
\begin{equation*}
  \bar{n} = Kn / p
\end{equation*}
for $p = \sum_{k} p_{k}$, to be interpreted as the average number of observed data points per factor level.
\end{definition}
The multigrid decomposition relates the rate of convergence of the Gibbs sampling Markov chain to that of simpler Markov chains. We summarise the main result in Proposition \ref{prop:multigrid} below, which is an amalgamation of Theorem 1, Lemmas 1 and 2 of \cite{papaspiliopoulos2020scalable}, extended to the case of local centering. The proof is analogous to the ones of the results just mentioned, and is omitted here for brevity.

\begin{proposition}[Multigrid decomposition] \label{prop:multigrid}
For a factor $k$, let $\bar{a}^{(k)} = \sum_{i} a_{i}^{(k)}/p_{k}$ be the factor average, and $\delta a_{i}^{(k)} = a_{i}^{(k)} - \bar{a}^{(k)}$ be the factor residuals. Let $\{\ba(t)\}$, where $t$ indexes iteration, be the Markov chain generated by either (GS), (cGS) or (Loc). Let $\{\bphi(t)\}$ and $\{\bpsi(t)\}$ be the processes that correspond to the factor averages (including $a^{(0)}$) and factor residuals respectively. Then, under balanced levels, the processes $\{\bphi(t)\}$ and $\{\bpsi(t)\}$ are Markov chains and independent of each other, and the rate of convergence of $\{\ba(t)\}$ is the maximum of their convergence rates. Additionally, the convergence rate of $\{\bpsi(t)\}$ is the same for all three algorithms.
\end{proposition}

Proposition \ref{prop:multigrid} reduces the computation of the rate of convergence of the algorithm of interest to that of the rates of two different Markov chains, one low-dimensional that operates on factor averages, and another high-dimensional on factor residuals. The latter is common for all algorithms considered, therefore the algorithms differ in the speed of the chain of the averages. Although defined on a space of fixed dimension, the chain of the averages can have a mixing time that deteriorates as $n$ and $p$ grow. In fact, Proposition 3 of \cite{papaspiliopoulos2020scalable} shows that vanilla Gibbs sampling is not scalable precisely for this reason in regimes where both $n$ and $p$ grow. On the other hand, the same proposition shows that the chain of averages generates independent samples in the case of the collapsed Gibbs sampler, therefore the rate of convergence of this algorithm is completely determined by that of the residuals. While in the case of local centering the chain of averages does not consist of independent draws, in Proposition \ref{prop:loc_c_averages} we show that it has a relaxation time that is bounded by constants that do not depend on $n$ and $p$.
The remaining results of this section, whose proofs can be found in Appendix \ref{app:proofs}, are derived under the following assumption:
\begin{enumerate}
\item[(A1)] Assume model \eqref{eq:crossed_Gauss1d} with balanced levels and an improper flat prior on $a^{(0)}$, i.e. $\mu_{pr} = \tau_{pr} = 0$.
\end{enumerate}

\begin{proposition}\label{prop:loc_c_averages}
Under (A1), the relaxation time of the Markov chain of averages $\{\bpsi(t)\}$ induced by (Loc) is upper bounded by $1 + \min_{k=1,\dots,K} \tau_{k} / \tau$.
\end{proposition}

The previous results imply that (cGS) and (Loc) have similar convergence properties, as formally shown below.

\begin{corollary}\label{thm:loc_c_equiv_to_collaps}
Under (A1), the relaxation times of (cGS) and (Loc), denoted by $T_{\mathrm{cgs}}$ and $T_{\mathrm{loc}}$ respectively, satisfy
\begin{equation*}
  T_{\mathrm{cgs}} \leq T_{\mathrm{loc}} \leq T_{\mathrm{cgs}} + C
\end{equation*}
where $C$ is a constant depending only on $\bg$ and not on $n$, $p$ or $K$.
\end{corollary}

Obviously, for non-Gaussian likelihoods for which the sampling of the $\xi_{i}^{(k)}$ cannot be done directly, but where a gradient-based sampler is used instead, one would expect a further increase in relaxation time. However, as shown by numerical results in the above sections, this will typically involve only constants that do not depend on $n$ and $p$.

Therefore, in view of Corollary \ref{thm:loc_c_equiv_to_collaps}, for the rest of the section we focus on the collapsed Gibbs sampler, and refer to its computational complexity as $\Cost(\mathrm{cGS})$. We now provide a more refined result on its scalability in the two-factor case, i.e.\ $K = 2$, building on previous work in \cite{papaspiliopoulos2020scalable}. On the basis of the previous discussion, a characterisation of the relaxation time of the chain of residuals is central to the following result. It involves an auxiliary Markov chain on a discrete state-space that we introduce first.

\begin{definition}\label{def:taux}
$T_{\mathrm{aux}}$ is the relaxation time of the two-component deterministic-scan Gibbs sampler that targets the discrete distribution with state space $\{1, \dots, p_{1}\} \times \{1, \dots, p_{2}\}$ and probability mass function proportional to $n_{ii'}^{(1,2)}$ for $(i,i')$ being an element of the state space.
\end{definition}

\begin{theorem}\label{prop:Gibbs_compl}
Assume (A1) and $K=2$. Then 
\begin{equation}\label{eq:cost_gibbs_bound}
\Cost(\mathrm{cGS}) = \bigo{\max\{n,p\} \min\{\bar{n}, T_{\mathrm{aux}}\}},
\end{equation}
where the constants depend only on $\bg$ and not on $n$ or $p$.
\end{theorem}

The factor $\max\{n,p\}$ comes from the cost per iteration, and $\min\{\bar{n}, T_{\mathrm{aux}}\}$ comes from the relaxation time, which in view of the previous discussion is that of the chain of residuals.

To obtain some intuition about the implications of this result, it is convenient to interpret the 2-factor ($K=2$) model as a recommender system, where one factor pertains to users, and the other to products. Note that $\bar{n}$ can be interpreted as the average number of products rated by a customer; it is precisely this number when $p_{1} = p_{2}$ and the design has balanced levels. In very sparse designs, increasing values of $n$ will not be associated with increasing values of $\bar{n}$ (more products and customers, but each customer rates a bounded number of products). In such regimes the sampler is scalable. In less sparse designs where $\bar{n}$ increases with $n$, the contingency table is more populated, and provided it is populated enough for the auxiliary Markov chain to mix well, the sampler will again be scalable.

The above discussion suggest that $\bar{n}$ and $T_{\mathrm{aux}}$ typically exhibit complementary behaviour with respect to sparsity of the data co-occurrence matrix. However, the value of $T_{\mathrm{aux}}$ does not only depend on the level of sparsity,  but also on the specific observation pattern, encoded in the co-occurrence counts $n^{(1,2)} = (n_{ii'}^{(1,2)})_{i=1,\dots,p_{1};i'=1,\dots,p_{2}}$, and can vary greatly across different designs with the same sparsity level. Indeed,  $T_{\mathrm{aux}}$ is the only term in the bound  \eqref{eq:cost_gibbs_bound} that does not only depend on the amount of data and parameters in the model, but also on the observation pattern. We now perform a more refined analysis of $\Cost(cGS)$ in \eqref{eq:cost_gibbs_bound} by considering random observation patterns and relating $T_{\mathrm{aux}}$ to the rich literature on spectral properties of random walks on random graphs.

We consider random binary designs with balanced levels. Specifically, given non-negative integers $d_{1}$, $d_{2}$ and $n$ such that $n$ is a multiple of both $d_{1}$ and $d_{2}$, we define $\mathcal{D}(n, d_{1}, d_{2})$ as the collection of all $p_{1}\times p_{2}$ binary designs with balanced levels and $n$ observations in total, where $p_{1}= n/d_{1}$, $p_{2} = n/d_{2}$. In other words $\mathcal{D}(n, d_{1}, d_{2})$ contains all designs $n^{(1,2)}$ such that $n_{i}^{(1)} = d_{1}$, $n_{i'}^{(2)} = d_{2}$ and $n_{ii'}^{(1,2)} \in \{0,1\}$ for all $i = 1, \dots, p_{1}$ and $i' = 1, \dots, p_{2}$. We will assume that the observed design is sampled uniformly from such collection, i.e.\ $n^{(1,2)}\sim \Unif(\mathcal{D}(n, d_{1} , d_{2}))$. The following theorem provides a characterization of $T_{\mathrm{aux}}$ and thus of $\Cost(cGS)$ in such regimes.

\begin{theorem}\label{thm:average_case}
Assume (A1) and $K = 2$. Let $d_{1}$ and $d_{2}$ be fixed integers larger than 4, and assume $n^{(1,2)} \sim \Unif(\mathcal{D}(n, d_{1}, d_{2}))$.
Then, for any $\epsilon > 0$,  we have
\begin{equation}\label{eq:alon_bound}
  T_{\mathrm{aux}} \leq 1 + \frac{2}{\sqrt{\min{\{d_{1}, d_{2}\}}} - 2} + \epsilon
\end{equation}
and
\begin{equation}\label{eq:cgs_scalable_on_average}
\Cost(cGS)= \bigo{\max\{n,p\}},
\end{equation}
asymptotically almost surely as $n\to\infty$, with constants in \eqref{eq:cgs_scalable_on_average} depending only on $\bg$ and not on $n$, $p$, $d_{1}$ or $d_{2}$.
\end{theorem}

Theorem \ref{thm:average_case} suggests that the collapsed Gibbs Sampler for 2-factor models is scalable in the average case, where average refers to the data co-occurrence pattern $n^{(1,2)}$. In particular, the almost sure statements \eqref{eq:alon_bound}-\eqref{eq:cgs_scalable_on_average} imply that for fixed $d_{1}$ and $d_{2}$, the fractions of balanced levels designs in $\mathcal{D}(n, d_{1}, d_{2})$ for which the collapsed Gibbs sampler is scalable goes to $1$ as $n \to \infty$.

As shown in the proof of Theorem \ref{thm:average_case}, which can be found in Appendix \ref{app:proofs}, \eqref{eq:alon_bound} follows from relating $T_{\mathrm{aux}}$ to the spectrum of random bipartite bi-regular graphs. This allows to then apply an extension for bipartite graphs, which was recently developed in \cite{brito2022spectral}, of the so-called Friedman's second largest eigenvalue theorem. The statement in \eqref{eq:cgs_scalable_on_average} then follows by combining \eqref{eq:alon_bound} and Theorem \ref{prop:Gibbs_compl}. Thus, at a high level, the positive result in \eqref{eq:cgs_scalable_on_average} is related to the fact that random graphs have near-optimal connectivity properties (i.e.\ they are expanders), as shown by a large body of random graph theory literature, see e.g.\ references in \cite{brito2022spectral}.

Interestingly, the upper bound in \eqref{eq:alon_bound} goes to $1$ as $d_{1}, d_{2} \to \infty$. This suggests that in regimes where $d_{1}$ and $d_{2}$ grow with $n$, which is a reasonable assumption in many contexts, then $T_{\mathrm{aux}}$ will get closer and closer to $1$, which in turn by Theorem \ref{prop:Gibbs_compl} implies that the relaxation time of the collapsed Gibbs Sampler will converge to $1$. While this cannot be deduced from Theorem \ref{thm:average_case}, since the regime studied there keeps $d_{1}$ and $d_{2}$ fixed and does not allow for $d_{1}$ and $d_{2}$ to grow $n$, the above results still point to a \emph{blessing of dimensionality}, where the sampling algorithm actually converges faster and faster as $n$ and $p$ grow. This conjecture is coherent with the empirical results in, e.g., Figure \ref{fig:crossed_lc_v_vanilla} as well as results in \cite{papaspiliopoulos2020scalable, ghosh2020backfitting}.

The theory developed in this section holds for Gaussian likelihoods. We see this more as a necessity of our strategy in proving the result, as opposed to the nature of the result itself, which we believe relates to the conditional independence structure in the model.  This is, again, corroborated by the findings in Figure \ref{fig:crossed_lc_v_vanilla}, which suggest that both the change from Gibbs to Metropolis-within-Gibbs, from known to unknown $\bg$ and from Gaussian to non-Gaussian likelihood slow down convergence only by a factor that is constant w.r.t.\ $n$ and $p$, and do not affect the scalability of the resulting MCMC algorithm.

\subsection{Related literature and alternative approaches}
\label{sec:crossed_disc}

Early work on Bayesian computation for crossed effects models includes \cite{vines1996fitting} and \cite[Section 6]{gelfand1996efficient}, who discuss, respectively, the use of identifiability constraints and reparameterisation for computational purposes. \cite{gao2017efficient,gao2020estimation} provide a systematic discussion of the super-linear cost of standard frequentist  fitting procedures when $K=2$, and develop a scalable method of moments. \cite{papaspiliopoulos2020scalable} provide a detailed analysis of the Gibbs sampler complexity, covering also sparse balanced designs and $K\geq 2$ factors, and propose the  collapsed Gibbs sampler for Gaussian likelihood. \cite{ghosh2020backfitting} propose a coordinate-wise descent method for computing the MAP estimator (phrased in their article as a backfitting procedure to compute generalized least squares estimates) that has close connections to collapsed Gibbs sampling. They prove $\bigo{\max\{n,p\}}$ complexity results also for somewhat unbalanced designs, their arguments are based on simple but effective concentration inequalities and linear algebra techniques.

A common practice among practitioners is the imposition of constraints within each factor, for example $\sum_{i} \ba_{i}^{(k)} = \bO$. Of course, such constraints are not needed for the inference to make sense in a Bayesian formulation, if desired they can be imposed a posteriori and different constraints will lead to different inferences.  To avoid confusion this is an entirely  different issue compared to the identifiability issues that arise specifically to categorical logistic models and are discussed in Section \ref{sec:crossed_multi}. From the point of view of this article it is worth considering their impact on algorithmic performance. Theorem 6 of \cite{zanella2020multilevel} derives the relaxation times of the vanilla Gibbs sampler for crossed effects models with Gaussian likelihood under identifiability constraints. Some constraints can make the vanilla Gibbs sampler scalable, whereas others cannot. Unfortunately, those found to be effective lead to difficult sampling problems when considering non-Gaussian likelihoods.

One could combine collapsing with techniques that seek to reduce posterior dependence between $\ba$ and $\bg$, such as parameter expansion \cite{liu1999parameter,meng1999seeking}. While useful and applicable, in the simulation studies of \cite{papaspiliopoulos2020scalable} for crossed effects models with Gaussian likelihoods, this strategy was found to have relatively limited impact on the resulting MCMC efficiency and it did not appear to change the overall complexity of the parent algorithm. Nevertheless, this is an important aspect of designing useful algorithms for large scale problems. 

\cite{menictas2019streamlined} and \cite{goplerud2022fast} propose mean field variational Bayes procedures for crossed models with Gaussian and logistic likelihoods, respectively, motivated by the overly high computational cost of Laplace approximations and HMC sampling for such models. There is an interesting interplay and scope for cross-fertilization between the methodology we develop in this article and variational inference and we will report our progress in that direction in future work.

Many popular software for models with Gaussian latent variables, such as \emph{lme4} \cite{bates2014fitting} or \emph{INLA} \cite{rue2009approximate}, perform approximate integration of $\ba$ using variants of the Laplace approximation. However, these approaches rely on sparse linear algebra techniques, which in general incurs super-linear complexity in $\max\{n,p\}$ for crossed effects models, see e.g.\ numerical results and related discussions in \cite{gao2017efficient,menictas2019streamlined,papaspiliopoulos2020scalable,gao2020estimation,ghosh2020backfitting,ghosh2022scalable}. On the contrary, such methods are entirely effective for nested multilevel models, as discussed in the following section.

\section{Nested multilevel models}
\label{sec:nested}

We consider the following broad class of nested multilevel hierarchical models. There is an underlying hierarchical organization in the data that is structured as a tree. To give a concrete example, in the large scale applications we consider later in this Section, the spatial domain in Spain is organized according to 5-digit postal codes where any longer sequence of digits (from 2 to 5) corresponds to an administrative unit (e.g., municipality) that is nested geographically within those that omit the last digits (e.g., city, province etc). A different example is classes, nested within schools, nested within municipalities, etc. We denote the root of the tree as $i_{0}$, its offsprings as $i_{0} i_{1}$, for $i_{1} = 1,\ldots,n_{0}$, for each such offspring its offsprings are $i_{0} i_{1} i_{2}$, for $i_{2} = 1,\ldots, n_{i_{0} i_{1}}$, etc. Overall there are $K$ levels apart from the root (hence $K=4$ in the postal code example). In full generality we observe a matrix of covariates $\bX_{i_{0}\cdots i_{k}}$ and a vector of responses $\by_{i_{0}\cdots i_{k}} $ and at every node in the tree, although it is more common that such observations are available only at level $K$. For example, in one of the real estate models we consider in the sequel for each leaf at $K=4$ the covariate is a  national house pricing index and the response local average house sale prices, with replications given over different quarters. The model relates the response to the covariates via a linear predictor:
\begin{equation*}
 \beeta_{i_{0}\cdots i_{k}}  = \bX_{i_{0}\cdots i_{k}} \bbb_{i_{0}\cdots i_{k}}.
\end{equation*}
Inference borrows strength across units by shrinking the regression parameters to those at the deeper level $\bbb_{i_{0}\cdots i_{k-1}}$ in terms of a Gaussian shrinkage prior:
\begin{equation*}
 \law(\bbb_{i_{0}\cdots i_{k}} \mid \bbb_{i_{0}\cdots i_{k-1}})  = \Nor({\bA_{i_{0}\cdots i_{k}}} \bbb_{i_{0}\cdots i_{k-1}}, \bT_{i_{0}\cdots i_{k-1}}^{-1}),
\end{equation*}
 where $\bA_{i_{0}\cdots i_{k}}$'s are  design matrices, often simply the identity matrices. Putting these ingredients together the full model is as follows: 
\begin{equation}\label{eq:nested_models}
\begin{aligned}
  \law(\bbb_{i_0}) & = \Nor(\bmu_{pr}, \bT_{pr}^{-1}), \\
  \law(\bT_{i_{0}\cdots i_{k}}) & = \Wish(\nu_{i_{0}\cdots i_{k}}, \bI / \nu_{i_{0}\cdots i_{k}}), \\
  \law(\bbb_{i_{0}\cdots i_{k}} \mid \bbb_{i_{0}\cdots i_{k-1}}, \bT_{i_{0}\cdots i_{k-1}}) & = \Nor({\bA_{i_{0}\cdots i_{k}}} \bbb_{i_{0}\cdots i_{k-1}}, \bT_{i_{0}\cdots i_{k-1}}^{-1}), \\
  \beeta_{i_{0}\cdots i_{k}} & = \bX_{i_{0}\cdots i_{k}} \bbb_{i_{0}\cdots i_{k}}, \\
  \law(\by_{i_{0}\cdots i_{k}} \mid \cdot ) & = \law(\by_{i_{0}\cdots i_{k}} \mid \beeta_{i_{0}\cdots i_{k}} ), \\
  k & = 0,\dots,K\,.
\end{aligned}
\end{equation}
With Gaussian likelihoods the posterior $\law(\bbb \mid \by, \bg)$ is multivariate Gaussian that can be sampled directly in a provably scalable way using sparse linear algebra. The result is given in Proposition \ref{prop:sla_nested}. For non-Gaussian likelihoods this result can be leveraged in different ways. One, which we explore in this work, is in situations where latent variables $\bz$ can be augmented such that $\law(\bbb \mid \by, \bz, \bg)$ is Gaussian. This is in the same spirit as the collapsed Gibbs sampling with Polya-gamma augmentation that we developed in Section \ref{sec:gibbs_polya} for crossed effects models. In Section \ref{sec:ngauss-bp} we show results for binomial and Cauchy likelihoods. A different strategy appropriate for light-tailed likelihoods, such as the Poisson, is to combine the efficient Gaussian sampling with distributed gradient-based sampling of the regression parameters on the leaves of the tree; this is also discussed in Section \ref{sec:ngauss-bp}.

\subsection{Methodology for Gaussian likelihoods: sparse linear algebra}
\label{sec:gauss-bp}

For nested multilevel models with Gaussian likelihoods, we can generate samples from the posterior $\law(\bbb \mid \by, \bg)$ directly at $\bigo{\max\{n,p\}}$ cost, as shown in Proposition \ref{prop:sla_nested}. Specifically, we consider the model \eqref{eq:nested_models} with likelihood
\begin{equation}
  \law(\by_{i_{0}\cdots i_{k}} \mid \beeta_{i_{0}\cdots i_{k}}) = \Nor(\beeta_{i_{0}\cdots i_{k}}, \tau_{i_{0}\cdots i_{k}}^{-1}\bI)\,.
\label{eq:gauss_lik_nested}
\end{equation}
As a consequence of \eqref{eq:nested_models}-\eqref{eq:gauss_lik_nested}, the conditional distribution $\law(\bbb \mid \by, \bg)$ is multivariate Gaussian. Sampling a multivariate Gaussian distribution can be reduced to solving linear systems; for details on this perspective and associated tools we refer to \cite{rue2005gaussian}, who provides results that can be used to prove the following claims. When, given appropriate matrices $\bV$, $\bU$ and $\bR$, the model can be expressed as
\begin{equation}
  \label{eq:gauss_post}
  \begin{aligned}
    \law(\bbb \mid \bg) & = \Nor(\bm, \bV^{-1}) \\
    -2 \log p(\by \mid \bbb, \bg) & = \bbb^{\mathsf{T}} \bU \bbb -2 \by^{\mathsf{T}} \bR^{\mathsf{T}} \bbb
  \end{aligned}
\end{equation}
(the second equation up to constants in $\bbb$), then $\law(\bbb \mid \by, \bg)$ is also a Gaussian distribution with precision $\bQ = \bU + \bV$. Let $\bL$ be the lower-triangular Cholesky factor of $\bQ$, and $\bz$ be a standard Gaussian vector. Then, a sample from the posterior $\law(\bbb \mid \by, \bg)$ is obtained by solving the following two systems:
\begin{equation}\label{eq:gauss-sim}
\begin{aligned}
  \bL \bw & = \bR \by + \bV \bm, \\
  \bL^{\mathsf{T}} \bbb & = \bw + \bz.
\end{aligned}
\end{equation}
It is well-known that $\bL$ can be computed column-wise from first to last column. Hence, the computation of $\bL$, and the simultaneous solution of the top equation in \eqref{eq:gauss-sim} by forward substitution, can be thought of as a \emph{forward pass}. Having completed the forward pass, we can then solve the second equation in \eqref{eq:gauss-sim} by backward substitution, in what we can think of as a \emph{backward pass}. The cost of solving these systems is dominated by the cost of computing $\bL$, which is in general cubic in the dimension of $\bbb$.

A direct calculation verifies the following claims. Let $\bQ$ denote the posterior precision associated with the Gaussian $\law(\bbb \mid \by,\bg)$. We refer to the elements of $\bQ$ by indicating what blocks of regression coefficients they pertain to, rather than labeling the blocks with ordinal indices. For instance, we write $\bQ[\bbb_{i_{0}\cdots i_{k}}, \bbb_{i_{0}\cdots i_{k-1}}]$ to refer to the sub-matrix of $\bQ$ that corresponds to $\bbb_{i_{0}\cdots i_{k}}$ and $\bbb_{i_{0}\cdots i_{k-1}}$ in the nested multilevel model. In the nested specification \eqref{eq:nested_models}, $\bbb_{i_{0}\cdots i_{k}}$ and $\bbb_{i_{0}\cdots i_{k'}}$ are conditionally independent unless $k' \in \{k-1, k, k+1\}$, hence $\bQ[\bbb_{i_{0}\cdots i_{k}}, \bbb_{i_{0}\cdots i_{k'}}] = \bO \bO^{\mathsf{T}}$. Accordingly, $\bQ$ is sparse in the sense that it only has $\bigo{p}$ non-zero elements. These are located in the following positions (with the obvious additional non-zero blocks due to symmetry):
\begin{equation}\label{eq:Q_nested}
\begin{aligned}
  \bQ[\bbb_{i_{0}\cdots i_{k}}, \bbb_{i_{0}\cdots i_{k-1}}]
  & = -\bT_{i_{0}\cdots i_{k-1}} \bA_{i_{0}\cdots i_{k}}, \\
  \bQ[\bbb_{i_{0}\cdots i_{k}}, \bbb_{i_{0}\cdots i_{k}}]
  & = \begin{cases} \begin{aligned} & \bT_{i_{0}\cdots i_{k-1}} + \tau_{i_{0}\cdots i_{k}} \bX_{i_{0}\cdots i_{k}}^{\mathsf{T}} \bX_{i_{0}\cdots i_{k}} \\
                                    & + n_{i_{0}\cdots i_{k}} \bA_{i_{0}\cdots i_{k+1}}^{\mathsf{T}} \bT_{i_{0}\cdots i_{k}} \bA_{i_{0}\cdots i_{k+1}} \end{aligned} & (k < K) \\
                    \bT_{i_{0}\cdots i_{k-1}} + \tau_{i_{0}\cdots i_{k}} \bX_{i_{0}\cdots i_{k}}^{\mathsf{T}} \bX_{i_{0}\cdots i_{k}} & (k = K) \end{cases},
\end{aligned}
\end{equation}
where we define $\bT_{i_{0}i_{-1}} = \bT_{pr}$ for $k = 0$, and we recall that $n_{i_{0}\cdots i_{k}}$ is the number of offsprings of node $\bbb_{i_{0}\cdots i_{k}}$ in the graphical model. With respect to \eqref{eq:nested_models}-\eqref{eq:gauss_post}, for nested multilevel models $\bV \bm$ is a vector of zeros except for the location that corresponds to $\bbb_{0}$, in which case it takes the value $\bT_{pr} \bmu_{pr}$. Thus, the top equation in \eqref{eq:gauss-sim} is simplified accordingly. Finally, as defined in \eqref{eq:gauss_post}, $\bR\by$ can trivially be computed at $\bigo{\max\{n,p\}}$ cost.

The Cholesky factor $\bL$ does not necessarily inherit the sparsity of $\bQ$, and it is known that the ordering of the indices plays a role in the sparsity of the Cholesky factor. This role can be understood through the concept of future sets, see for example in Theorem 2.8 of \cite{rue2005gaussian}. Therefore, the computational performance of linear algebra methods for sampling from Gaussian distributions depends strongly on the way $\bbb$ and $\bQ$ are organized. We now define a specific ordering that turns out to be optimal for nested multilevel models.

\begin{definition}
\label{def:depth}
For nested multilevel models as defined in \eqref{eq:nested_models}, we define the ``depth-last'' ordering, according to which $\bbb$ is organized by placing $\bbb_{0}$ last, then the $\bbb_{i_{0} i_{1}}$'s (the order among them is irrelevant), then the $\bbb_{i_{0} i_{1} i_{2}}$'s, and so on, until we reach the level furthest from the root.
\end{definition}

For nested multilevel models, Proposition \ref{prop:sla_nested} shows that under the ``depth-last'' ordering the Cholesky factor $\bL$ has the same sparsity as $\bQ$, i.e. it has $\bigo{p}$ non-zero elements, and that it may be computed at $\bigo{p}$ cost. Therefore, \eqref{eq:gauss-sim} can be used at a $\bigo{\max\{n,p\}}$ cost to draw samples from $p(\bbb \mid \by, \bg)$. The proposition shows how the non-zero elements can be computed sequentially. However, the main consequence of the Proposition is to establish the sparsity of $\bL$ under the given ordering. Any sparse linear algebra software (e.g. CHOLMOD/SuiteSparse) will efficiently compute $\bL$ from a representation of $\bQ$ as a sparse matrix with ``depth-last'' ordering, hence the user would not need to hard-code the iterations described in the proposition.

\begin{proposition}
  \label{prop:sla_nested}
  For the nested multilevel model as defined in \eqref{eq:nested_models} and \eqref{eq:gauss_lik_nested}, and under the ``depth-last'' ordering defined in Definition \ref{def:depth}, let $\bQ$ be the corresponding posterior precision matrix and $\bL$ its associated Cholesky factor. Then, $\bL$ has the same sparsity structure as $\bQ$ and its only non-zero blocks are computed according to
  \begin{equation}\label{eq:bp-chol}
    \begin{aligned}
      \bL[\bbb_{i_{0}\cdots i_{k}}, \bbb_{i_{0}\cdots i_{k}}]^{\mathsf{T}} \bL[\bbb_{i_{0}\cdots i_{k}}, \bbb_{i_{0}\cdots i_{k-1}}]
      & = -\bT_{i_{0}\cdots i_{k-1}} \bA_{i_{0}\cdots i_{k}}, \\
      \bL[\bbb_{i_{0}\cdots i_{k}}, \bbb_{i_{0}\cdots i_{k}}] \bL[\bbb_{i_{0}\cdots i_{k}}, \bbb_{i_{0}\cdots i_{k}}]^{\mathsf{T}}
      & = \bC_{i_{0}\cdots i_{k}} + \bT_{i_{0}\cdots i_{k-1}},
    \end{aligned}
  \end{equation}
  where as before $\bT_{i_{0}\cdots i_{k-1}} = \bT_{pr}$ for $k = 0$. The matrices $\bC_{i_{0}\cdots i_{k}}$ are defined according to the recursion
  \begin{equation}\label{eq:post-root}
    \begin{aligned}
      \bC_{i_{0}\cdots i_{k}} & = \begin{cases} \tau_{i_{0}\cdots i_{K}} \bX_{i_{0}\cdots i_{K}}^{\mathsf{T}} \bX_{i_{0}\cdots i_{K}} & (k = K) \\
                                    \sum_{j_{k+1}} \tilde{\bC}_{i_{0}\cdots i_{k} j_{k+1}} + \tau_{i_{0}\cdots i_{k}} \bX_{i_{0}\cdots i_{k}}^{\mathsf{T}} \bX_{i_{0}\cdots i_{k}}& (k < K) \end{cases}, \\
      \tilde{\bC}_{i_{0}\cdots i_{k}} & = \bA_{i_{0}\cdots i_{k}}^{\mathsf{T}} \bB_{i_{0}\cdots i_{k}}^{\mathsf{T}} (\bB_{i_{0}\cdots i_{k}} \bT_{i_{0}\cdots i_{k-1}}^{-1} \bB_{i_{0}\cdots i_{k}}^{\mathsf{T}} + \bI)^{-1} \bB_{i_{0}\cdots i_{k}} \bA_{i_{0}\cdots i_{k}},
    \end{aligned}
  \end{equation}
  where $\bB_{i_{0}\cdots i_{k}}$ follows from the decomposition $\bC_{i_{0}\cdots i_{k}} = \bB_{i_{0}\cdots i_{k}}^{\mathsf{T}} \bB_{i_{0}\cdots i_{k}}$, and the notation $\sum_{j_{k+1}} \tilde{\bC}_{i_{0}\cdots i_{k} j_{k+1}}$ is to be understood as summing over the children of node $i_{0}\cdots i_{k}$. 
\end{proposition}

The cost of carrying out the recursion in Proposition \ref{prop:sla_nested} is dominated by $\bigo{p}$ decompositions of $L \times L$-dimensional matrices, where recall that $L$ is the dimension of the regression parameters at each node. Therefore, the non-zero elements are obtained at $\bigo{pL^{3}}$ cost. The Proposition can be proved in different ways, one using a future set argument, and another using \emph{belief propagation}. In fact, the sequence of computational steps induced by sparse linear algebra based on ``depth-last'' ordering is almost identical to the one induced by belief propagation algorithms applied to the joint distributions defined by the posterior of the nested multilevel model. Belief propagation, as well as the general junction tree algorithm, is a framework for computations in graphical models, see Chapter 6 of \cite{cowell2007probabilistic} and Chapter 2 of \cite{wainwright2008graphical}, but for nested multilevel models the two paradigms are essentially equivalent. For an introduction to belief propagation see Section 8.4 in \cite{bishop2006pattern}. For details on how belief propagation can be used for nested multilevel models see the technical report \cite{papaspiliopoulos2017note}, which also provides a historical perspective on this and links to various references. For other references discussing scalable Sparse Linear Algebra (SLA) methods to perform matrix factorisation for sparse matrices with non-zero elements corresponding to a tree see e.g.\ \cite[Section 13.2]{vishnoi2013lx} and references therein. Given that the proof can be worked out on the basis of one of these established approaches, we omit it from the present article.

\begin{figure}[h]
  \centering
  \includegraphics{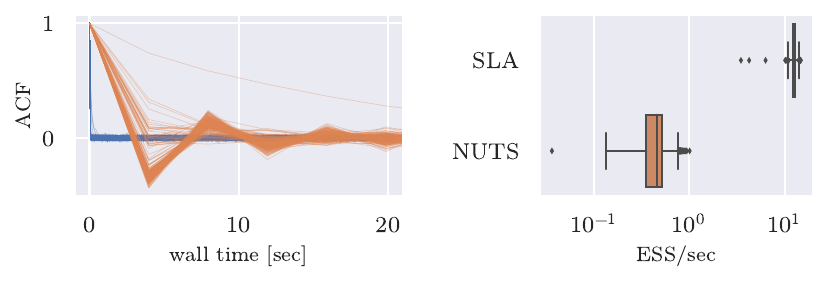}
  \caption{[\emph{Nested effects real estate model, Gaussian likelihood}] Comparison of sampling efficiency between MCMC based on Gibbs with SLA (blue) and Stan/NUTS (orange). The left panel overlays ACFs of $\bbb$ and $\bg$ as a function of wall time. The right panel shows the distribution of effective samples per second. Due to the large number of parameters, we only plot ACFs for a subset. For each algorithm, we include ACFs for the 20 slowest parameters and a representative subsample of the rest. Estimates are computed from MCMC runs with 10000 (SLA) and 1000 iterations (NUTS) respectively, with an warm-up period of 1000 iterations for NUTS. We set the hyperparameters $\nu_{k} = L$ for $\bT_{k}$, a flat prior on $\bbb_{0}$, and $\tau \sim \Gam(1/2, 1/2)$. \label{fig:nested_gaussian_benchmark}}
\end{figure}

\subsection{Methodology for non-Gaussian likelihoods: data augmentation and blocked Gibbs sampling}
\label{sec:ngauss-bp}

The direct sampling approach based on belief propagation or sparse linear algebra can still be useful to build scalable sampling algorithms even when the likelihoods are non-Gaussian. One such setting arises when conditionally on latent variables $\bz_{i_{0}\cdots i_{k}}$ associated to each observation $\by_{i_{0}\cdots i_{k}}$, $\law(\bbb \mid \by, \bz, \bg)$ is Gaussian. We have already discussed this approach with the binomial likelihood in Section \ref{sec:gibbs_polya}, and we can transfer the result to the nested setting. Moreover, heavier-tailed likelihoods, e.g. the Laplace or the Student-t, can be represented as scale mixtures of Gaussians, i.e.
\begin{equation}
\begin{aligned}
  & \law(\by_{i_{0}\cdots i_{k}} \mid \beeta_{i_{0}\cdots i_{k}}) \\
  & \qquad = \int_{0}^{\infty} \Nor(\by_{i_{0}\cdots i_{k}}; \beeta_{i_{0}\cdots i_{k}}, (z_{i_{0}\cdots i_{k}} / \tau_{i_{0}\cdots i_{k}})\bI) p(z_{i_{0}\cdots i_{k}}) dz_{i_{0}\cdots i_{k}}.
\end{aligned}
\end{equation}
Letting $\bz$ denote the vector of all such latent variables, our proposed scheme is to work with the data augmented target $\law(\bbb, \bg, \bz \mid \by)$, and target it with a Gibbs sampler, where we sample from $\law(\bbb \mid \by, \bg, \bz)$ using the methodology of Section \ref{sec:gauss-bp}. $\law(\bz \mid \by, \bbb, \bg)$ decomposes according to
\begin{equation}
  \law(\bz \mid \by, \bbb, \bg) = \prod_{k=0}^{K} \prod_{i_{0}\cdots i_{k}} \law(z_{i_{0}\cdots i_{k}} \mid \by_{i_{0}\cdots i_{k}}, \bbb, \bg),
\end{equation}
with each $z_{i_{0}\cdots i_{k}}$ easily updated.

Various important specifications fall outside of this class, such as the Poisson regression model $\law(y_{i_{0}\cdots i_{k}} \mid \eta_{i_{0}\cdots i_{k}}) = \Pois(e^{\eta_{i_{0}\cdots i_{k}}})$. This instance is most easily described when the data $\by$ is only observed at the leaf level $K$, which in any case is the most common. Defining $\bbb^{(K)}$ as the set of all regression parameters $\bbb_{i_{0}\cdots i_{K}}$ at level $K$, we suggest the strategy of alternating updates to $\law(\bbb^{(-K)} \mid \bbb^{(K)}, \bg)$ and $\law(\bbb^{(K)} \mid \by, \bbb^{(K-1)}, \bg)$, noting that $\bbb^{(-K)}$ is conditionally independent of $\by$. Indeed, $\bbb^{(-K)}$ is now structured as a tree with $K-1$ levels and Gaussian data $\bbb^{(K)}$ at the new leaves. Thus, we sample from the full conditional distribution law of $\bbb^{(-K)}$ according to the methodology of Section \ref{sec:gauss-bp}. Conversely, while $\law(\bbb^{(K)} \mid \by, \bbb^{(K-1)}, \bg)$ is not Gaussian, it decomposes according to
\begin{equation}
  \law(\bbb^{(K)} \mid \by, \bbb^{(K-1)}, \bg) = \prod_{i_{0}\cdots i_{K}} \law(\bbb_{i_{0}\cdots i_{K}} \mid \by_{i_{0}\cdots i_{K}}, \bbb^{(K-1)}, \bg),
\end{equation}
with each $\bbb_{i_{0}\cdots i_{K}}$ updated independently by a Metropolis-within-Gibbs step akin to the one presented in Section \ref{sec:gibbs_method_nGauss}, at a total cost of order $\bigo{p}$ for the full update from $\law(\bbb^{(K)} \mid \by, \bbb^{(K-1)}, \bg)$. While the cost of the combined update remains $\bigo{p}$, the main potential weakness of that strategy lies in the induced dependence between $\bbb^{(K)}$ and $\bbb^{(-K)}$, which in some settings increases the relaxation time of the algorithm as $p$ increases. The work in \cite{papaspiliopoulos2007general} and the associated theory in \cite{papaspiliopoulos2008stability} suggests that for likelihood with tails lighter than the Gaussian, such as the Poisson, this centered parameterisation (in the terminology of \cite{papaspiliopoulos2007general}) is expected to perform well, whereas for heavy-tailed likelihoods it would not. Hence, we view the two strategies proposed in this Section as complementary.

\begin{figure}[h]
  \centering
  \includegraphics{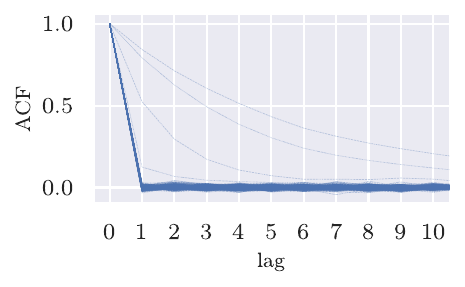}
  \caption{[\emph{Nested effects real estate model, binomial likelihood}] Sampling efficiency of the PG/SLA algorithm for binomial likelihoods. The plot overlays ACFs of $\bbb$ and $\bg$ as a function of MCMC lag. Due to the large number of parameters, we only plot ACFs for a subset: we include ACFs for the 20 slowest parameters and a representative subsample of the rest. Estimates are computed from an MCMC run with 10000 iterations. We set the hyperparameters $\nu_{k} = L$ for $\bT_{k}$ and ($\bmu_{pr} = \bO$, $\bT_{pr} = \bI$) for $\bbb_{0}$, corresponding to a standard Gaussian prior. \label{fig:nested_binomial_sareb}}
\end{figure}

\begin{figure}[h]
  \centering
  \includegraphics{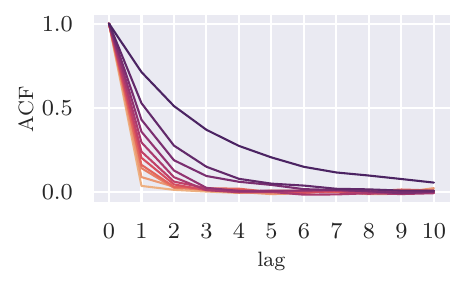}
  \caption{[\emph{Nested effects real estate model, Cauchy likelihood}] Comparison of sampling efficiency of the GMix/SLAwG algorithm for different proportions $\alpha = 0.1, 0.2, \ldots, 1$ of heavy-tailed observation densities. The panel shows ACFs of the model likelihood for proportions ranging from $\alpha=0$ (lightest) to $\alpha=1$ (darkest). Estimates are computed from an MCMC run with 10000 iterations. We set the hyperparameters $\nu_{k} = L$ for $\bT_{k}$, a flat prior on $\bbb_{0}$, and $\tau \sim \Gam(1/2, 1/2)$. \label{fig:nested_heavyt}}
\end{figure}

\begin{figure}[h]
  \centering
  \includegraphics{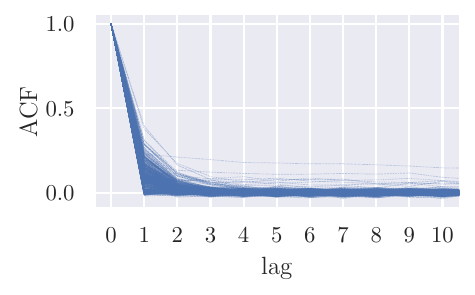}
  \caption{[\emph{Nested effects real estate model, Poisson likelihood}] Sampling efficiency of the LT/SLAwG algorithm for Poisson likelihoods. The plot overlays ACFs of $\bbb$ and $\bg$ as a function of MCMC lag. Due to the large number of parameters, we only plot ACFs for a subset: we include ACFs for the 20 slowest parameters and a representative subsample of the rest. Estimates are computed from an MCMC run with 10000 iterations. We set the hyperparameters $\nu_{k} = L$ for $\bT_{k}$ and ($\bmu_{pr} = \bO$, $\bT_{pr} = \bI$) for $\bbb_{0}$, corresponding to a standard Gaussian prior. \label{fig:nested_poisson_sareb}}
\end{figure}

\subsection{Numerical experiments with a real estate model}
\label{sec:nested_data}

We assess algorithmic performance on a semi-synthetic dataset that originates from a broader task of temporal prediction of real estate prices at high spatial resolution. Our approach combines a type of multilevel capital asset pricing model, which relates the local average square meter price of different type of real estate properties to national economic indices (e.g. GDP), with a multivariate time series model for these indices. In this article we focus on the multilevel regression component of this model. The spatial domain is Spain, organized according to 5-digit postal codes which define $K=4$ levels, plus the root. The first two digits define a province in the country and the later codes correspond to higher spatial resolutions. There are 44 observations at each postal code, each of which is the difference in consecutive quarters of the logarithm of local average sales price. The quarters span the period 2007-2018. The predictors, an intercept and a national housing price index (hence $L=2$), are only available at the leaf-level. In this data structure, there are $p \approx 8 \times 10^{3}$ regression parameters and $n \approx 4 \times 10^{6}$ observations. Due to confidentiality constraints, we work with a dataset that has been simulated from a misspecified version of \eqref{eq:nested_models}, using the part of the dataset that is publicly available and parameter values $\bbb$ that are consistent with the real data.

In two further experiments, we adopt the latent variable augmentation setting of \ref{sec:ngauss-bp}, but keep the same prior structure as in the Gaussian experiment. In one instance, we transform the response into a binary variable according to its sign, set $\law(y_{i_{0}\cdots i_{K} t} | \eta_{i_{0}\cdots i_{K} t}) = \Bin(1, (1+\exp\{-\eta_{i_{0}\cdots i_{K} t}\})^{-1})$, and use Polya-gamma augmentation to obtain conditionally Gaussian observations, as in Section \ref{sec:gibbs_polya}. We show results in Figure \ref{fig:nested_binomial_sareb}. Remarkably, the elements of $\bbb$ appear to mix instantly, while elements of $\bg$ have significantly larger IATs. These are ordered by the level that the parameter pertains to: the higher the level, the slower the mixing. This is due to the rising number of coefficients per hyperparameter, and therefore increasing strength of the prior. This could be remedied, e.g. by way of parameter expansion. In the other instance, we postulate a heterogeneous observation model where $\law(y_{i_{0}\cdots i_{K} t} \mid \eta_{i_{0}\cdots i_{K} t}) = \Cau(\eta_{i_{0}\cdots i_{K} t}, \tau^{-1/2})$ for a proportion $\alpha$ of the terminal leaves, and $\Nor(\eta_{i_{0}\cdots i_{K} t}, \tau^{-1})$ otherwise. Here, the $t$ subscript denotes replications over times for $t=1,\dots,44$. We vary $\alpha = 0.1, 0.2, \ldots, 1$, with $\alpha=0$ corresponding to the model and associated algorithm that is analyzed in Figures \ref{fig:nested_gaussian_benchmark}. For observations $y_{i_{0}\cdots i_{K} t}$ with non-Gaussian observation density we augment with latent variables $z_{i_{0}\cdots i_{K} t}$, given which the density is Gaussian, and proceed with the blocked Gibbs sampler outlined above for each of the resulting 10 models. In Figure \ref{fig:nested_heavyt} we plot the estimated ACF for the likelihood of the data (marginalizing over the latent variables $\bz$, where present). The experiment shows that as the proportion of latent variables increases the mixing of the algorithm slows down, but the deterioration is bounded. In particular, the autocorrelation is almost negligible after 10 iterations in all settings. Considering the large value of $n$ and $p$ of this dataset, the results are coherent with the conjecture that the algorithm's relaxation time remains small as $n$ and $p$ diverge.

In a final experiment, the response $y_{i_{0}\cdots i_{K} t}$ is the number of transactions at the location $i_{0}\cdots i_{K}$ and time $t$, and we model it as $\Pois(\exp\{\eta_{i_{0}\cdots i_{K} t}\})$. There is a trend in the transaction numbers, which we account for by introducing the time index $t$ as an additional covariate, and therefore $L=3$. We apply our strategy for light-tailed likelihoods as laid out in Section \ref{sec:ngauss-bp}, with updates to $\bbb_{i_{0}\cdots i_{K}}$ proposed according to a second order expansion of $\law(\bbb_{i_{0}\cdots i_{K}} \mid \by_{i_{0}\cdots i_{K}}, \bbb^{(K-1)}, \bg)$. We show results in Figure \ref{fig:nested_poisson_sareb}. The algorithm achieves good mixing throughout the tree, with the exception of a few nodes located on the penultimate level $L=3$. Our investigation suggests that this is due to the fact that the model we use for the data is purposely mis-specified and this creates the slower, although by no means slow, mixing of a few internal nodes. In fact, the mis-specification creates a non-trivial dependence between precision hyperparameters and a few regression parameters.

\section{Discussion}
\label{sec:beyond}

We have developed efficient sampling algorithms for large scale Bayesian inference with two of the most popular families of models for applied research, crossed effects and nested multilevel models with Gaussian priors and non-Gaussian likelihoods, and we have obtained rigorous results on their scalability properties. Our results support the claim that appropriately designed blocked Gibbs samplers can effectively exploit the sparse conditional dependence structure inherent to these models in order to achieve an overall computational complexity that is $\bigo{\max\{n, p\}}$. From a practical point of view, these algorithms are observed to provide dramatic computational speed-ups compared to off-the-shelf HMC, both for simulated data and in challenging applications.

There are various open questions and directions for future research that would complement and extend the present work, two of which we elaborate on. Firstly, we worked under the assumption that scalable sampling from the conditional distribution of the high-dimensional regression parameters, i.e.\ $\law(\ba \mid \by, \bg)$ and $\law(\bbb \mid \by, \bg)$ for crossed and nested models respectively, is sufficient to scalably sample from the joint distributions $\law(\ba, \bg \mid \by)$ and $\law(\bbb, \bg \mid \by)$, at least in commonly encountered scenarios. While this is strongly supported by simulations, theoretical results on when the assumption is met, and when it isn't, would further strengthen our conclusions. A first step in this direction is made in \cite{ascolani2023complexity}, which analyses the two-block Gibbs sampler targeting the joint distribution $\law(\bbb, \bg \mid \by)$ for two-level nested hierarchical models. The theory developed therein is asymptotic and supports the claim that under appropriate assumptions, the convergence time of the above two block sampler remains bounded as $p,n\to\infty$. Secondly, while we demonstrated that SLA methods are ideally suited to nested models, it turns out that they are not effective for crossed effect models. In ongoing work we are performing a systematic exploration of SLA for crossed models, both to substantiate this claim, as well as to assess the effectiveness of \emph{approximate} SLA methods, which sample from an approximation to $\law(\bbb \mid \by, \bg)$.

\section*{Acknowledgements}
 
The authors are most grateful for their inputs Mr Jie Hao Kwa, whose master dissertation in 2018 obtained useful insights into sparse linear algebra methods for nested multilevel models, and Maximilian M\"uller, whose master dissertation in 2020 investigated sparse linear methods for crossed effect models. The first two authors would like to thank Jose Garcia-Montalvo for the collaboration on the applied projects that have motivated this work. The article has benefited from comments by Darren Wilkinson. The third author acknowledges support from the European Research Council (ERC), through StG ``PrSc-HDBayLe'' grant ID 101076564.

\bibliography{bibliography}

\appendix

\section{Proofs}
\label{app:proofs}

\begin{proof}[Proof of Theorem \ref{prop:Gibbs_compl}] The cost per iteration has been established to be $\bigo{2n + p}=\bigo{\max\{n,p\}}$, hence the proof focuses on quantifying the relaxation time. Denote the relaxation time of (cGS) by $T_{\mathrm{cgs}}$. We now prove that
\begin{equation}\label{eq:relax_order}
  c \min\{\bar{n}, T_{\mathrm{aux}}\} \leq T_{\mathrm{cgs}} \leq C \min\{\bar{n}, T_{\mathrm{aux}}\}
\end{equation}
for any balanced level design and for some positive and finite constants $c$ and $C$, which depend only on $\bg$. The thesis follows then directly from \eqref{eq:relax_order}.

By \cite[Theorem 4]{papaspiliopoulos2020scalable} we have $T_{\mathrm{cgs}} = (1 - \rho_{1}\rho_{2}\rho_{\mathrm{aux}})^{-1}$ where $\rho_{k} = (n\tau + p_{k}\tau_{k})^{-1} n\tau$ for $k = 1,2$; $\rho_{\mathrm{aux}}$ is the $L^2$-rate of convergence of the auxiliary Gibbs sampler satisfying $T_{\mathrm{aux}} = (1 - \rho_{\mathrm{aux}})^{-1}$, $\tau$ is the likelihood precision and $\tau_{k}$ the prior precision of the $k$-th factor. Since $\rho \mapsto (1 - \rho)^{-1}$ is increasing on $(0,1)$ and $\rho_{1}\rho_{2}\rho_{\mathrm{aux}} \leq \min\{\rho_{1}, \rho_{2}, \rho_{\mathrm{aux}}\}$, it follows that
\begin{equation*}
  T_{\mathrm{cgs}} \leq (1 - \min\{\rho_{1}, \rho_{2}, \rho_{\mathrm{aux}}\})^{-1}  = \min\left\{(1-\min\{\rho_{1}, \rho_{2}\})^{-1}, (1-\rho_{\mathrm{aux}})^{-1}\right\}.
\end{equation*}
By definition of $\rho_{1}$ and $\rho_{2}$,
\begin{equation}\label{eq:upper_rho_12}
\begin{aligned}
  (1 - \min\{\rho_{1}, \rho_{2}\})^{-1} & = 1 + \frac{n\tau}{\max\{p_{1}\tau_{1}, p_{2}\tau_{2}\}} \\
  & \leq 1 + C_{1} \frac{n}{\max\{p_{1}, p_{2}\}} \\
  & \leq 1 + C_{1}\bar{n}, 
\end{aligned}
\end{equation}
with $C_{1} = \frac{\tau}{\min\{\tau_{1}, \tau_{2}\}}$ and exploiting $\max\{p_{1}, p_{2}\} \geq 2^{-1}(p_{1} + p_{2})$. Combining the above inequalities with $T_{\mathrm{aux}} = (1 - \rho_{\mathrm{aux}})^{-1}$ we obtain
\begin{equation*}
  T_{\mathrm{cgs}}
    \leq \min\left\{1 + C_{1}\bar{n}, T_{\mathrm{aux}}\right\}
    \leq C \min\left\{\bar{n}, T_{\mathrm{aux}}\right\},
\end{equation*}
where we can take $C = 1 + C_{1}$ since $\bar{n}\geq 1$ by construction.

For the lower bound, $\rho_{1}\rho_{2}\rho_{\mathrm{aux}}\geq(\min\{\rho_{1}, \rho_{2}, \rho_{\mathrm{aux}}\})^{3}$ implies
\begin{equation*}
\begin{aligned}
  T_{\mathrm{cgs}} & \geq (1-\min\{\rho_{1}, \rho_{2}, \rho_{\mathrm{aux}}\}^3)^{-1} \\
  & \geq 3^{-1}(1 - \min\{\rho_{1}, \rho_{2}, \rho_{\mathrm{aux}}\})^{-1},
\end{aligned}
\end{equation*}
where the second inequality follows from $(1 - \rho^t)^{-1} \geq t^{-1} (1 - \rho)^{-1}$ for every $t \geq 1$ and $\rho \in (0,1)$. To prove the latter claim note that $(1 - \rho^t)^{-1} \geq t^{-1}(1 - \rho)^{-1}$ if and only if $\rho^t - t\rho + (t - 1) \geq 0$ and that the left-hand side is a decreasing function of $\rho$ on $(0, 1)$ that vanishes when $\rho = 0$. It follows that
\begin{equation*}
  T_{\mathrm{cgs}} \geq 3^{-1} \min \left\{(1 - \min\{\rho_{1}, \rho_{2}\})^{-1}, (1 - \rho_{\mathrm{aux}})^{-1}\right\}.
\end{equation*}
Then, analogously to \eqref{eq:upper_rho_12},
\begin{equation*}
\begin{aligned}
  (1 - \min\{\rho_{1}, \rho_{2}\})^{-1} & \geq \frac{n\tau}{\max\{p_{1}\tau_{1}, p_{2}\tau_{2}\}} \\
  & \geq c_{1} \frac{n}{\max\{p_{1}, p_{2}\}} \\
  & \geq \frac{c_{1}}{2}\bar{n},
\end{aligned}
\end{equation*}
with $c_{1} = \frac{\tau}{\max\{\tau_{1}, \tau_{2}\}}$ and exploiting $\max\{p_{1}, p_{2}\} \leq p_{1} + p_{2}$. 
Thus, we obtain
\begin{equation*}
  T_{\mathrm{cgs}}
    \geq \min\left\{\frac{c_{1}}{2}\bar{n}, T_{\mathrm{aux}}\right\}
    \geq c \min\left\{\bar{n}, T_{\mathrm{aux}}\right\},
\end{equation*}
where $c = \min\{1, c_{1}/2\}$ as desired.
\end{proof}

\begin{proof}[Proof of Proposition \ref{prop:loc_c_averages}]
Throughout the proof, $\{\bar{\ba}(t)\}$ denotes the chain of averages. We highlight that $a^{(0)}$ takes $K+1$ intermediate values, denoted by $(a^{(0)}(t, k))_{k=0}^{K}$.
The chain evolves according to the following scheme.
\begin{algorithmic}
\State Given $\bar{\ba}(t) = (a^{(0)}(t), \bar{a}^{(1)}(t), \dots, \bar{a}^{(K)}(t))$ do:
\State Set $a^{(0)}(t, 0) \gets a^{(0)}(t)$
  \For{$k = 1:K$}
  \State Draw $a^{(0)}(t, k)$ from
  \begin{equation*}
    \Nor\left(a^{(0)}(t, k-1) + \bar{a}^{(k)}(t), (p_{k}\tau_{k})^{-1}\right)
  \end{equation*}
  \State Draw $\bar{a}^{(k)}(t + 1)$ from
  \begin{equation*}
    \Nor\begin{pmatrix}\rho_{k} \left(\bar{y} - a^{(0)}(t, k) - \sum_{\ell=1}^{k-1} \bar{a}^{(\ell)}(t + 1) - \sum_{\ell=k+1}^{K} \bar{a}^{(\ell)}(t)
\right), \\ (n\tau + p_{k}\tau_{k})^{-1}\end{pmatrix}
  \end{equation*}
  \EndFor
  \State Set $a^{(0)}(t + 1) \gets a^{(0)}(t, K)$
\end{algorithmic}
By construction
\begin{equation*}
\begin{aligned}
  \mathbb{E}[a^{(0)}(t,k) | \bar{\ba}(t)] & = \mathbb{E}[\mathbb{E}[a^{(0)}(t, k)|a^{(0)}(t, k-1), \bar{\ba}(t)] | \bar{\ba}(t)] \\
  & = \mathbb{E}[a^{(0)}(t, k-1) | \bar{\ba}(t)] + \bar{a}^{(k)}(t),
\end{aligned}
\end{equation*}
which by induction over $k$ implies that $\mathbb{E}[a^{(0)}(t, k) | \bar{\ba}(t)] = {a}^{(0)}(t) + \sum_{\ell=1}^k \bar{a}^{(\ell)}(t)$ for $k = 1, \dots, K$ and thus $\mathbb{E}[a^{(0)}(t + 1) | \bar{\ba}(t)] = {a}^{(0)}(t) + \sum_{\ell=1}^{K} \bar{a}^{(\ell)}(t)$. Moreover, for $k = 1, \dots, K$
\begin{equation}\label{eq:iter_ind}
\begin{aligned}
  \mathbb{E}[\bar{a}^{(k)}(t) | \bar{\ba}(t)] & = \rho_{k} (\bar{y} - \mathbb{E}[a^{(0)}(t, k) | \bar{\ba}(t)] + \textstyle\sum_{\ell=k+1}^{K} \bar{a}^{(\ell)}(t) \\
  & \qquad + \textstyle\sum_{\ell=1}^{k-1} \mathbb{E}[\bar{a}^{(\ell)}(t + 1) | \bar{\ba}(t)]) \\
  & = -\rho_{k} (a^{(0)}(t) + \textstyle\sum_{\ell=1}^{K} \bar{a}^{(\ell)}(t) \\
  & \qquad + \textstyle\sum_{\ell=1}^{k-1} \mathbb{E}[\bar{a}^{(\ell)}(t + 1) | \bar{\ba}(t)]) + \mathrm{const},
\end{aligned}
\end{equation}
where we use ``$\mathrm{const}$'' to indicate fixed values that do now depend on $\bar{\ba}(t)$. It follows from \eqref{eq:iter_ind} that $\mathbb{E}[\bar{a}^{(1)}(t) | \bar{\ba}(t)] = -\rho_{1} \left(a^{(0)}(t) + \sum_{\ell=1}^{K} \bar{a}^{(\ell)}(t)\right) + \mathrm{const}$ and, by induction over $k = 1, \dots, K$,
\begin{equation*}
  \mathbb{E}[\bar{a}^{(k)}(t) | \bar{\ba}(t)] = c_{k} \left(a^{(0)}(t) + \sum_{\ell=1}^{K} \bar{a}^{(\ell)}(t)\right) + \mathrm{const},
\end{equation*}
with $(c_{k})_{k=1}^{K}$ satisfying $c_{k} = -\rho_{k}(1 + \sum_{\ell=1}^{k-1}c_{\ell})$. Therefore, the transition from $\bar{\ba}(t)$ to $\bar{\ba}(t+1)$ corresponds to a linear autoregressive process with a $(K+1) \times (K+1)$ autoregressive matrix $\bB$. By well-known theory for Gaussian autoregression processes, the rate of convergence of such a Markov chain equals the largest modulus eigenvalue of the matrix $\bB$, which we denote by $\rho(\bB)$. Defining $c_{0} = 1$, the above results imply that $B_{k\ell} = c_{k}$ for all $k, \ell = 0, \dots, K$, meaning that the $K+1$ columns of $B$ are all equal to $(c_{0}, \dots, c_{K})^{\mathrm{T}}$. It follows that $B$ is a rank $1$ matrix with $K$ eigenvalues equal to 0 and one non-zero eigenvalue equal to $\sum_{k=0}^{K} c_{k}$. Hence, to bound $\rho(\bB)$ we need to upper bound the absolute value of $\sum_{k=0}^{K} c_{k}$. Define $d_{k} = \sum_{\ell=1}^{k} c_{\ell}$ for $k = 1, \dots, K$. We now prove by induction that: $d_{k} \in (-1, 0)$, $d_{k} \leq -\rho_{k}$ and $d_{k}$ is monotonously decreasing in $k = 1, \dots, K$. The hypothesis is true for $k=1$ since $d_{1} = -\rho_{1}$ and $\rho_{k} \in (0,1)$ for all $k = 1, \dots, K$. For $k \geq 2$, we have
\begin{equation*}
\begin{aligned}
  d_{k} & = \sum_{\ell=1}^k c_\ell = d_{k-1} + c_{k} = d_{k-1} - \rho_{k}(1 + d_{k-1}) \\
  & = (1 - \rho_{k}) d_{k-1} - \rho_{k}.
\end{aligned}
\end{equation*}
Since $(1 - \rho_{k}) \in (0,1)$ and $d_{k-1} \in (-1, 0)$, it follows that $(1 - \rho_{k}) d_{k-1} \in (-1, 0)$ and $d_{k} \leq -\rho_{k}$. Also, being a convex combination of $d_{k-1} \in (-1, 0)$ and of $-1$, it follows $d_{k} \in (-1, 0)$ and $d_{k} < d_{k-1}$. This proves the induction hypothesis. Since $d_{K} \in (-1, 0)$ we have $\sum_{k=0}^{K} c_{k} = 1 + d_{K} > 0$. By monotonicity of $d_{k}$ and $d_{k} \leq -\rho_{k}$ for $k = 1, \dots, K$, it follows that $d_{k} \leq \min_{k=1,\dots,K} -\rho_{k}$. Thus $\sum_{k=0}^K c_{k} = 1 + d_{K}\leq \min_{k=1,\dots,K} (1 - \rho_{k})$. Accordingly,
\begin{equation*}
  \rho(\bB) = \sum_{k=0}^{K} c_{k} \leq \min_{k=1,\dots,K} 1 - \rho_{k} = \min_{k=1,\dots,K} \frac{p_{k}\tau_{k}}{n\tau + p_{k}\tau_{k}}.
\end{equation*}
It follows that the relaxation time of the $(\bar{a}(t))_{t\geq 1}$ chain equals
\begin{equation*}
\begin{aligned}
  (1 - \rho(\bB))^{-1} & \leq \left(1 - \min_{k=1,\dots,K} \frac{p_{k} \tau_{k}}{n\tau + p_{k}\tau_{k}}\right)^{-1} \\
  & \quad = 1 + \min_{k=1,\dots,K} \frac{p_{k}\tau_{k}}{n\tau} \\
  & \quad \leq 1 + \frac{\tau_{\text{min}}}{\tau} = \bigo{1},
\end{aligned}
\end{equation*}
where $\tau_{\text{min}} = \min_{k=1,\dots,K} \tau_{k}$ is a fixed constant in our analysis. Above we used $p_{k} \leq n$ for all $k = 1, \dots, K$, which holds by construction as every level of every factor gets at least one observation.
\end{proof}

\begin{proof}[Proof of Theorem \ref{thm:average_case}]
By the argument in the proof of \cite[Prop.4]{papaspiliopoulos2020scalable}, we obtain that
\begin{equation}\label{eq:t_aux_spectr}
T_{\mathrm{aux}} = (1 - \lambda_{2}^{*}(P_{1} P_{2}))^{-1},
\end{equation}
where $P_{1} = n^{(1,2)} / d_{1}$ and $P_{2} = (n^{(1,2)})^{\mathrm{T}} / d_{2}$ are, respectively, $p_{1} \times p_{2}$ and $p_{2} \times p_{1}$ stochastic matrices; and $\lambda_{2}^{*}(M)$ denotes the second largest modulus eigenvalue of a matrix $M$. The spectrum of the $p_{1} \times p_{1}$ transition matrix $P_{1} P_{2}$ is related to the one of the $(p_{1} + p_{2}) \times (p_{1} + p_{2})$ matrix
\begin{equation}\label{eq:def_A}
  A = \left(
    \begin{array}{c|c}
      \bO_{p_{1} \times p_{2}} & n^{(1,2)} \\
      \hline
      (n^{(1,2)})^{\mathrm{T}} & \bO_{p_{2} \times p_{1}}
    \end{array}
  \right),
\end{equation}
where $\bO_{m \times r}$ denotes a zero matrix of dimension $m \times r$. The matrix $A$ can be interpreted as the adjacency matrix of a bipartite graph with $p_{1} + p_{2}$ nodes,  one for each parameter in $\ba^{(1)}$ and $\ba^{(2)}$, and edges between $a_{i}^{(1)}$ and $a_{i'}^{(2)}$ if and only if $n_{ii'}^{(1,2)} = 1$. The definitions of $A$ and $P_{1} P_{2}$ imply that
\begin{equation}\label{eq:conn_design_graph}
\lambda_{2}^{*}(P_{1} P_{2}) = \lambda_{2}(A) / \sqrt{d_{1} d_{2}},
\end{equation}
where $\lambda_{2}(M)$ denotes the second largest eigenvalue of a matrix $M$\footnote{More generally the whole spectrum of $P_{1} P_{2}$ and $A$ are related: assuming $p_{1} < p_{2}$ without loss of generality, if $\lambda$ is an eigenvalue of $P_{1} P_{2}$, then $\sqrt{d_{1} d_{2}} \lambda$ and $-\sqrt{d_{1} d_{2}} \lambda$ are eigenvalues of $A$, and the remaining $p_{2} - p_{1}$ eigenvalues of $A$ are zero. See e.g. \cite{brito2022spectral} for details.}.

The bipartite graph associated to $A$, denoted by $G(A)$, is also bi-regular, or more specifically $(d_{1}, d_{2})$-regular, meaning that the degree of all nodes in the $k$-th component is $d_{k}$, for $k \in \{1, 2\}$. Moreover, it can be seen that the map from $n^{(1,2)}$ to $G(A)$, with $A$ is as in \eqref{eq:def_A}, defines a bijection from $\mathcal{D}(n, d_{1}, d_{2})$ to $\mathcal{G}(n, d_{1}, d_{2})$, where the latter denotes the collection of all $(d_{1}, d_{2})$-regular bipartite graphs with a total of $n$ edges. It follows that, when $n^{(1,2)} \sim \Unif(\mathcal{D}(n, d_{1}, d_{2}))$ and $A$ is as in \eqref{eq:def_A}, then $G(A) \sim \Unif(\mathcal{G}(n, d_{1}, d_{2}))$. Thus, by \eqref{eq:conn_design_graph}, we need to study the second largest eigenvalue of the adjacency matrix of a random bipartite bi-regular graph. This has been studied in \cite{brito2022spectral}. Theorem 3.2 therein implies that $\lambda_{2}(A) \leq \sqrt{d_{1}-1}+\sqrt{d_{2}-1}+\epsilon'$ almost surely as $n \to \infty$ for any $\epsilon' > 0$, for fixed $d_{1}$ and $d_{2}$ larger or equal than $3$.\footnote{The condition $d_{1}, d_{2}\geq 3$ is assumed at the beginning of \cite{brito2022spectral} and not explicitly recalled in Theorem 3.2.}
Combining the latter with \eqref{eq:t_aux_spectr} and \eqref{eq:conn_design_graph} and performing some basic manipulations we obtain that, for every $\epsilon>0$, it holds almost surely as $n\to\infty$  that
\begin{align*}
  T_{\mathrm{aux}}
  & \leq \frac{\sqrt{d_{1} d_{2}}}{\sqrt{d_{1} d_{2}} - \sqrt{d_{1}-1} - \sqrt{d_{2} - 1}} + \epsilon \\
  & \leq 1 + \frac{\sqrt{d_{1} - 1} + \sqrt{d_{2} - 1}}{\sqrt{d_{1} d_{2}} - \sqrt{d_{1} - 1} - \sqrt{d_{2} - 1}} + \epsilon \\
  & \leq 1 + \frac{2\sqrt{\max\{d_{1}, d_{2}\}}}{\sqrt{d_{1} d_{2}} - 2 \sqrt{\max\{d_{1}, d_{2}\}}} + \epsilon = 1 + \frac{2}{\sqrt{\min\{d_{1}, d_{2}\}} - 2} + \epsilon,
\end{align*}
where we used $d_{1}, d_{2}> 4$ to ensure that $\sqrt{\min\{d_{1}, d_{2}\}}-2>0$. The condition $d_{1}, d_{2}> 4$ is purely assumed for simplicity in order to allow for a more concise and interpretable bound on $T_{\mathrm{aux}}$.
This proves \eqref{eq:alon_bound}. Equation \eqref{eq:cgs_scalable_on_average} then follows directly combining \eqref{eq:cost_gibbs_bound} with $T_{\mathrm{aux}} \leq 1 + \frac{2}{\sqrt{\min\{d_{1}, d_{2}\}} - 2} \leq 1 + \frac{2}{\sqrt{5} - 2} < \infty$.
\end{proof}

\section{MCMC performance metrics}
\label{app:mcmc}

The \emph{integrated autocorrelation time} (IAT), is defined as
\begin{equation}
  \operatorname{iat} X = 1 + 2\sum_{t=1}^{\infty} \operatorname{acf}_{t} X,
\end{equation}
where $X$ is a stochastic process and $\operatorname{acf}_{t} X$ is the autocorrelation function (ACF) of $X$ evaluated at $t$-th lag. In practice, we estimate ACF by empirical averages from a single trajectory of $X$. We then use these estimates to estimate the IAT following Sokal's truncation heuristic \cite{sokal1997monte}. The heuristic consists of computing a running sum over ACF lags and stopping as soon as the running sum exceeds a multiple of the next lag. We modify this heuristic slightly and only truncate at even lags to account for NUTS' tendency to produce estimates of ACF of different sign at successive lags. We define the \emph{effective sample size} (ESS) as the run length divided by the IAT estimate. Since we are comparing methods with vastly different computing time per iteration, we monitor the run time of the algorithm for both algorithms and use \emph{effective sample size per second} as our performance metric. Notice that we include transience time in the ESS/sec estimate. Moreover, when overlaying ACFs for both methods, we plot them as a function of computing time, as opposed to iteration lag.

\section{Table of Algorithms}
\label{sec:table_alg}

\begin{table}[h!]
\centering
\begin{tabular}{llll}
\toprule
Acronym & Structure & Summary & Appears In \\
\midrule
LC-MwG & Crossed & Local centering & Figures \ref{fig:crossed_lc_v_vanilla}, \ref{fig:crossed_binomial_elections} \\
Van-MwG & Crossed & Vanilla Metropolis-within-Gibbs & Figure \ref{fig:crossed_lc_v_vanilla} \\
const/LC-MwG & Crossed & Local centering w/ ident. constr. & Figures \ref{fig:crossed_multinomial_benchmark}, \ref{fig:crossed_multinomial_ident} \\
const/Van-MwG & Crossed & Vanilla Gibbs w/ ident. constr. & Figure \ref{fig:crossed_multinomial_benchmark} \\
free/LC-MwG & Crossed & Local centering w/o ident. constr. & Figure \ref{fig:crossed_multinomial_ident} \\
free/PLC-MwG & Crossed & Proj. local centering w/o ident. constr. & Figure \ref{fig:crossed_multinomial_ident} \\
PG/Van-G & Crossed & Polya-gamma augm. vanilla Gibbs & Figure \ref{fig:crossed_binomial_elections} \\
PG/Col-G & Crossed & Polya-gamma augm. collapsed Gibbs & Figure \ref{fig:crossed_binomial_elections} \\
\midrule
SLA & Nested & Gaussian sparse lin. alg. & Figure \ref{fig:nested_gaussian_benchmark} \\
PG/SLAwG & Nested & Polya-gamma augm. sparse lin. alg. & Figure \ref{fig:nested_binomial_sareb} \\
GMix/SLAwG & Nested & Gaussian mixture sparse lin. alg. & Figure \ref{fig:nested_heavyt} \\
LT/SLAwG & Nested & Light-tailed likelihood sparse lin. alg. & Figure \ref{fig:nested_poisson_sareb} \\
\bottomrule
\end{tabular}
\caption{Algorithms and corresponding acronyms. \label{tb:acro}}
\end{table}

\end{document}